%% file: main.tex
\documentclass[10pt,journal,compsoc]{IEEEtran}

\usepackage{multirow}

\usepackage[hidelinks]{hyperref}

\usepackage{amsmath} 

\usepackage{enumitem}
\setlist[itemize]{noitemsep, topsep=0pt, leftmargin=*}

\usepackage[nopostdot,nogroupskip,nonumberlist]{glossaries}
\input{list_notations} 
\input{list_abbreviations}

\usepackage{balance}

\usepackage{cite}
\usepackage{array}
\usepackage{array}
\newcolumntype{P}[1]{>{\centering\arraybackslash}p{#1}}

\usepackage{graphicx}

\usepackage{xcolor} 
\definecolor{brightmaroon}{rgb}{0.6, 0.13, 0.28}

\begin{document}
%

\title{EAPS: Edge-Assisted Predictive Sleep \\Scheduling for 802.11 IoT Stations}


%
%
%

\author{\IEEEauthorblockN{Jaykumar Sheth\IEEEauthorrefmark{1}, Cyrus Miremadi\IEEEauthorrefmark{1}, Amir Dezfouli\IEEEauthorrefmark{2}, and 
Behnam Dezfouli\IEEEauthorrefmark{1}}\\
\IEEEauthorblockA{\small\IEEEauthorrefmark{1}Internet of Things Research Lab, Department of Computer Science and Engineering, Santa Clara University, USA
\\
\texttt{\{jsheth, cmiremadi, bdezfouli\}@scu.edu}\\
}
\IEEEauthorblockA{\small\IEEEauthorrefmark{2}Commonwealth Scientific and Industrial Research Organisation (CSIRO), Sydney, Australia
\\
\texttt{\{amir.dezfouli@data61.csiro.au\}}
}
}

\maketitle
\thispagestyle{plain}
\pagestyle{plain}
\begin{abstract}
The broad deployment of 802.11 (a.k.a., WiFi) access points and significant enhancement of the energy efficiency of these wireless transceivers has resulted in increasing interest in building 802.11-based IoT systems.
Unfortunately, the main energy efficiency mechanisms of 802.11, namely PSM and APSD, fall short when used in IoT applications.
PSM increases latency and intensifies channel access contention after each beacon instance, and APSD does not inform stations about when they need to wake up to receive their downlink packets.
In this paper, we present a new mechanism---\textit{edge-assisted predictive sleep scheduling} (EAPS)---to adjust the sleep duration of stations while they expect downlink packets.
We first implement a Linux-based access point that enables us to collect parameters affecting communication latency.
Using this access point, we build a testbed that, in addition to offering traffic pattern customization, replicates the characteristics of real-world environments.
We then use multiple machine learning algorithms to predict downlink packet delivery.
Our empirical evaluations confirm that when using EAPS the energy consumption of IoT stations is as low as PSM, whereas the delay of packet delivery is close to the case where the station is always awake.

\end{abstract}


%
\IEEEpeerreviewmaketitle

\section{Introduction}
\label{intro}
Nowadays, in addition to regular user devices such as phones, laptops, and tablets, many IoT devices such as security cameras, smart locks, and medical devices rely on the 802.11 standard.
Several reasons support the importance of this standard for IoT applications:
First, compared to cellular communication, this standard offers a high bandwidth over unlicensed bands.
In addition to reducing costs, these features are particularly useful in domains such as video surveillance, industrial control, and medical monitoring, where high bandwidth is necessary.
Second, 802.11 base stations---known as \glspl{AP}---are broadly deployed in various environments and provide a ready-to-use infrastructure for IoT connectivity.
For example, most Internet service providers offer WiFi-enabled modems, and companies such as Comcast and AT\&T implement large-scale hotspot areas in cities across the US \cite{jha2016exploring}.
Third, the power consumption of 802.11 {stations} has been considerably reduced during the past decade by introducing various power-save mechanisms as well as developing enhanced low-power RF transceiver technologies such as power gating and clock gating \cite{cypressb,BCM,tozlu2012wi}.
Accordingly, nowadays we can find more 802.11-based IoT devices in the market such as \textit{ring} security camera and doorbell \cite{ring_camera}, \textit{nest} security camera \cite{nest}, \textit{Schlage} locks \cite{schlage}, and \textit{LIFX} light bulbs \cite{lifx}, to mention a few.
Nevertheless, the increasing number of users and applications is stressing the capabilities of these networks to address application demands.

The 802.11 standard offers multiple power-saving mechanisms to support energy-constrained stations.
\gls{PSM} enables the stations to wake up periodically and check if the \gls{AP} has any buffered packet(s) for them.
The \gls{AP} periodically broadcasts beacon packets at certain intervals called \gls{BI} to inform the stations about their buffered packets.
Stations send PS-Poll packet to the \gls{AP} to request delivery.
\gls{PSM} significantly increases communication delay because stations can only receive downlink packets after each beacon instance.
The delay problem further exacerbates with the concurrent transmission of {PS-Poll} packets and the accumulation of downlink delivery after each beacon instance \cite{he2009analysis,manweiler2011avoiding}.
To address this concern, \gls{APSM} requires a station waiting for downlink packets to stay awake for a particular \textit{tail time} duration (e.g., 10 ms \cite{cypressb}) after each packet exchange with the \gls{AP} \cite{jang2016adaptive}.
The tail time may impose idle listening and further waste of energy, especially if the delay between uplink and downlink delivery is longer than tail time.
Another enhancement of PSM is the \gls{APSD}, which enables the stations to request downlink packet delivery by sending NULL uplink packets to the \gls{AP} \cite{vinhas2013performance}.
Nonetheless, since downlink delivery delay depends on a variety of factors, employing this mechanism is challenging.
A new power-saving mechanism introduced in 802.11ax is \gls{TWT}, which allows the station to set wake up agreements with the \gls{AP}. 
However, \gls{TWT} is only a local agreement with the \gls{MAC} layer and it does not specify when the downlink will be ready for the station to wake up.
In general, despite the numerous features offered by these power-saving mechanisms, the existing solutions and implementations focus merely on the energy-delay trade-off of regular user stations such as smartphones \cite{peck2015practical,jang2011snooze,rozner2010napman,pyles2012sapsm}, and no attention has been paid to IoT application scenarios.

Many IoT applications require the transmission of uplink reports by station and reception of commands from a remote server.
For example, consider a sample medical application where an IoT device reports an event and expects to receive actuation commands in return.
Another example is a security camera that transfers images whenever a motion is detected, and waits for a command to stream video if a particular object is recognized.
After the transmission of uplink packets, the IoT station has five options before receiving downlink packets:
\begin{itemize}
    \item (i) stay in awake mode---this is known as \gls{CAM}, 
    \item (ii) return to sleep mode and wake up during the next beacon period---\gls{PSM},
    \item (iii) stay in awake mode for a short time duration---\gls{APSM}'s tail time),
    \item (iv) send a Null packet to the \gls{AP} periodically to check if the downlink packet has arrived---\gls{APSD},
    \item  (v) wake up when the downlink packet is about to be delivered.
\end{itemize}
Case (i) minimizes delay but does not offer any power efficiency.
Case (ii) causes long end-to-end delays \cite{rozner2010napman,pyles2012sapsm} because the station has to wait until the next beacon instance, even if the actual round-trip delay is less than the time remaining until the next beacon.
Besides, the delay considerably increases when the station and server need to complete multiple rounds of packet exchange to make a decision\footnote{Not only for IoT applications, studies show that every 10 ms increase in network access results in a 1000 ms increase in page load time \cite{sundaresan2013web}.}.
Also, accumulating packet deliveries after each beacon results in intensified channel access contention and extends the delay of downlink delivery.
Case (iii) is only effective if the delivery delay is short; otherwise, this case results in power waste.
Case (iv) results in periodic wake ups and unnecessarily increases channel access contention.
Therefore, none of these cases are suitable for applications where both delay and energy efficiency are the essential performance metrics.
Applying case (v) \textit{requires an accurate estimation of the delay between uplink and downlink packets}, which is composed of the following delay components:
First, the uplink packet received over the wireless interface must be sent over the wired interface.
The second component is the interval between the instance the packet leaves the \gls{AP} until a response is received from the server.
Third, once the reply is received, the packet must compete with other downlink packets and be delivered to the station when it is in awake mode.
In this paper, we show that computing the third delay component is particularly challenging because it depends on various factors, including airtime utilization, the intensity of uplink and downlink communication, access category of packets, and wireless link quality.
This is also verified by the recent studies that show the delay experienced at \gls{AP} is more than 60\% of the total communication delay between a station and server \cite{pei2016wifi,hoiland2015good}.
Besides, the buffering mechanism employed in the Linux kernel's network layer and wireless driver further complicates the modeling and prediction of these delay components \cite{hoiland2017ending,showail2016buffer,hoiland2015good,heusse2003performance}.
For example, Intel's IWL and Qualcomm's ath9k and ath10 drivers perform packet scheduling; however, these algorithms are heuristically designed by vendors---further complicating delay estimations.

In this paper, we propose a novel mechanism called \textit{edge-assisted predictive sleep scheduling} (EAPS) to reduce the idle listening time and energy of stations when waiting for downlink packets.
At a high level, EAPS works as follows:
Once an uplink packet is received from an IoT station, the delivery delay is computed using machine learning techniques.
The estimated delay is then conveyed to the station using a high-priority data-plane queue.
The station then switches into sleep mode and wakes up at the scheduled time to retrieve downlink packet.

This paper introduces the following {contributions}:
\textit{First}, we present the implementation of a Linux-based \gls{AP} with new and modified kernel and user-space modules to keep track of the system operation in terms of parameters such as incoming and outgoing transmissions, buffer status, and interference level.
\textit{Second}, using the modified \gls{AP}, we build a configurable testbed that allows us to generate various traffic patterns similar to real-world deployments.
We characterize the similarities by introducing multiple metrics.
\textit{Third}, the collected information is then fed into a user-space module, which estimates the delay components.
We focus on wired-to-wireless switching delays and their prediction using various machine learning algorithms under different traffic scenarios.
EAPS runs at the network edge to ensure quick decision-making about sleep schedules, which implies that all the necessary computations to calculate sleep schedules are performed by \gls{AP} to avoid increasing the computational load of resource-constrained IoT devices.
\textit{Fourth}, we perform an empirical evaluation of the impact of delay prediction on both energy efficiency and timeliness in scenarios where IoT stations communicate with cloud and edge servers.
In terms of delay, EAPS outperforms PSM by 45\% in the cloud scenario and by 84\% in the edge scenario.
The energy consumption of EAPS is 26\% lower in the cloud scenario and 6\% in the edge scenario, compared to PSM.
In the edge scenario, the delay of EAPS is close to that of APSM, while its energy efficiency is 37\% higher.
In the cloud scenario, EAPS improves delay and energy efficiency by 41\% and 46\%, respectively, compared to APSM.

The rest of this paper is organized as follows.
We present delay components and implementation details of the AP in Section \ref{delay_analysis}. 
We present the edge-assisted sleep scheduling mechanism in Section \ref{Predictive Scheduling}, and empirical performance evaluations in Section \ref{perf_eval}. 
We overview the related works in Section \ref{related_work}, and finally
Section \ref{conclusion} concludes the paper.

\begin{table}[]
\setlength\extrarowheight{1.7pt}
\caption{Summary of key notations}
\footnotesize
\begin{tabular}{l|l}
Notation & Meaning \\\hline
$\delta_{a}$ & wireless to wired interface switching delay \\
$\delta_{b}$ & \gls{AP}-server communication delay (round trip) \\
$\delta_{c}$ & wire to wireless interface switching delay \\
${q} $ & qdisc queue utilization \\ 
$\widehat{{q}} $ & MAC queue utilization \\ 
$ {c}_{\mathrm{u}} $ & Channel utilization \\
$ {c}_{\mathrm{n}} $ & Channel noise \\
$ {w} $ & Retransmission count \\
$ {r}_{\mathrm{in}}^{\mathrm{wired}} $ & Input rate of wired interface \\
$p_i$            &   packet\\
$\delta(p_{i})$          &   Actual deadline of packet $p_{i}$ \\
$\delta^{\prime}(p_{i})$          &   Estimated deadline of packet $p_{i}$ \\
\end{tabular}
\label{notations}
\end{table}

\section{Delay Analysis and \gls{AP} Development}
\label{delay_analysis}
In this section, we first analyze the delay components between uplink and downlink packet delivery, and then present our proposed \gls{AP} architecture and implementation.
The proposed \gls{AP} enables us to collect features that are necessary for delay prediction and schedule dissemination to IoT stations.

\subsection{Delay Components}
\begin{figure}
\centering
  \includegraphics[width=0.8\linewidth]{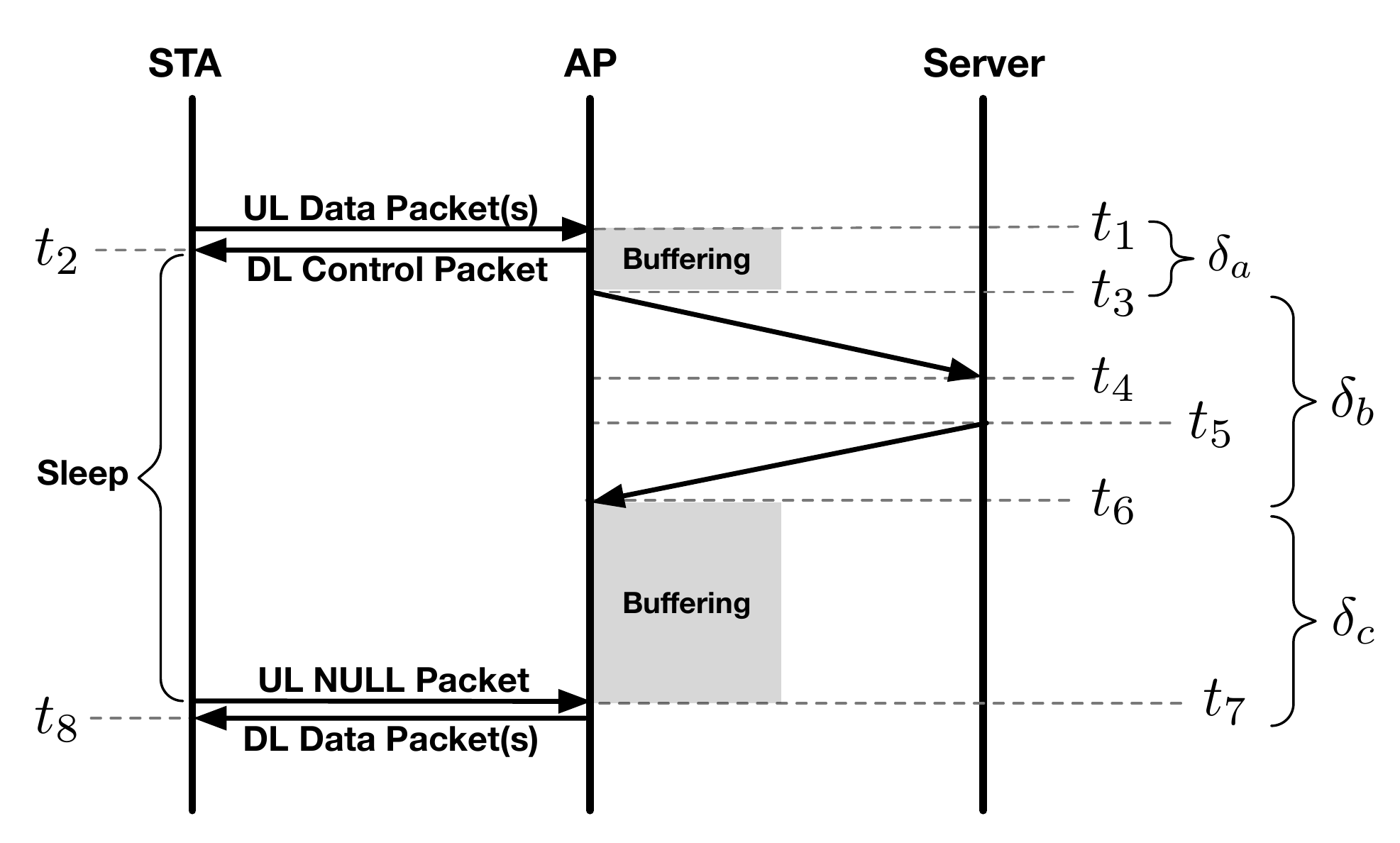}
  \caption{The end-to-end delay components between a station and a server. The prediction of $\delta_{c}$ is particularly challenging because it is affected by several factors such as traffic rate, channel utilization, and buffering mechanisms employed by the Linux kernel's network layer and wireless NIC's driver.}
  \label{fig:delays1}
\end{figure}

As Figure \ref{fig:delays1} shows, at time $t_{1}$ the station grasps the channel and transmits its uplink packet.
This uplink packet may represent a single uplink packet sent by the station or the last packet of a burst of uplink packets.
After this, the station waits to receive the downlink packet(s) from the \gls{AP}. 
We refer to the process of uplink and downlink packet exchange as \textit{transaction}.
In event-driven applications, the downlink packet is usually a command message issued by a server in response to the message sent by the station.
After the transmission of the uplink packet, the station has five options: 
(i) \gls{CAM}: stay awake until the reception of the downlink packet, 
(ii) \gls{PSM}: wake up at the next beacon instance, 
(iii) \gls{APSM}: stay awake for a short tail time duration, transition into sleep mode if no communication happens during the tail time, and wake up at the next beacon instance to receive the buffered packet,
(iv) \gls{APSD}: send a Null packet to the \gls{AP} periodically to check if the downlink packet has arrived,
and
(v) transition into sleep mode and wake up when the downlink packet is ready to be sent by the \gls{AP}. 
As explained in Section \ref{intro}, cases (i) through (iv) are not suitable for applications where both delay and energy efficiency are the essential performance metrics.
Therefore, in this paper, \textit{we focus on the case (v)} to enable the station to switch to sleep mode while waiting for downlink transmission.
To reduce the waiting time for downlink packet delivery, the station transitions into sleep mode after the reception of a control packet at $t_{2}$ and wakes up at $t_{7}$ to request and receive the downlink packet.
The sleep duration is conveyed to the station by the \gls{AP} through a control packet sent at $t_{2}$.
Therefore, we need to estimate the delay between $t_{1}$ to $t_{7}$.
To this end, we first modify a Linux-based \gls{AP} by adding new user-space and kernel-space modules.

It is worth noting that the uplink packet sent at $t_{7}$ may be followed by the delivery of multiple downlink packets.
As per the 802.11e amendment, stations can start a \gls{TXOP} for downlink delivery.
For example, if the traffic belongs to the voice \gls{AC}, the \gls{AP} will reserve a 1.504 ms slot for downlink delivery to the station.


\subsection{AP Development}
\begin{figure}
\centering
  \includegraphics[width=1\linewidth]{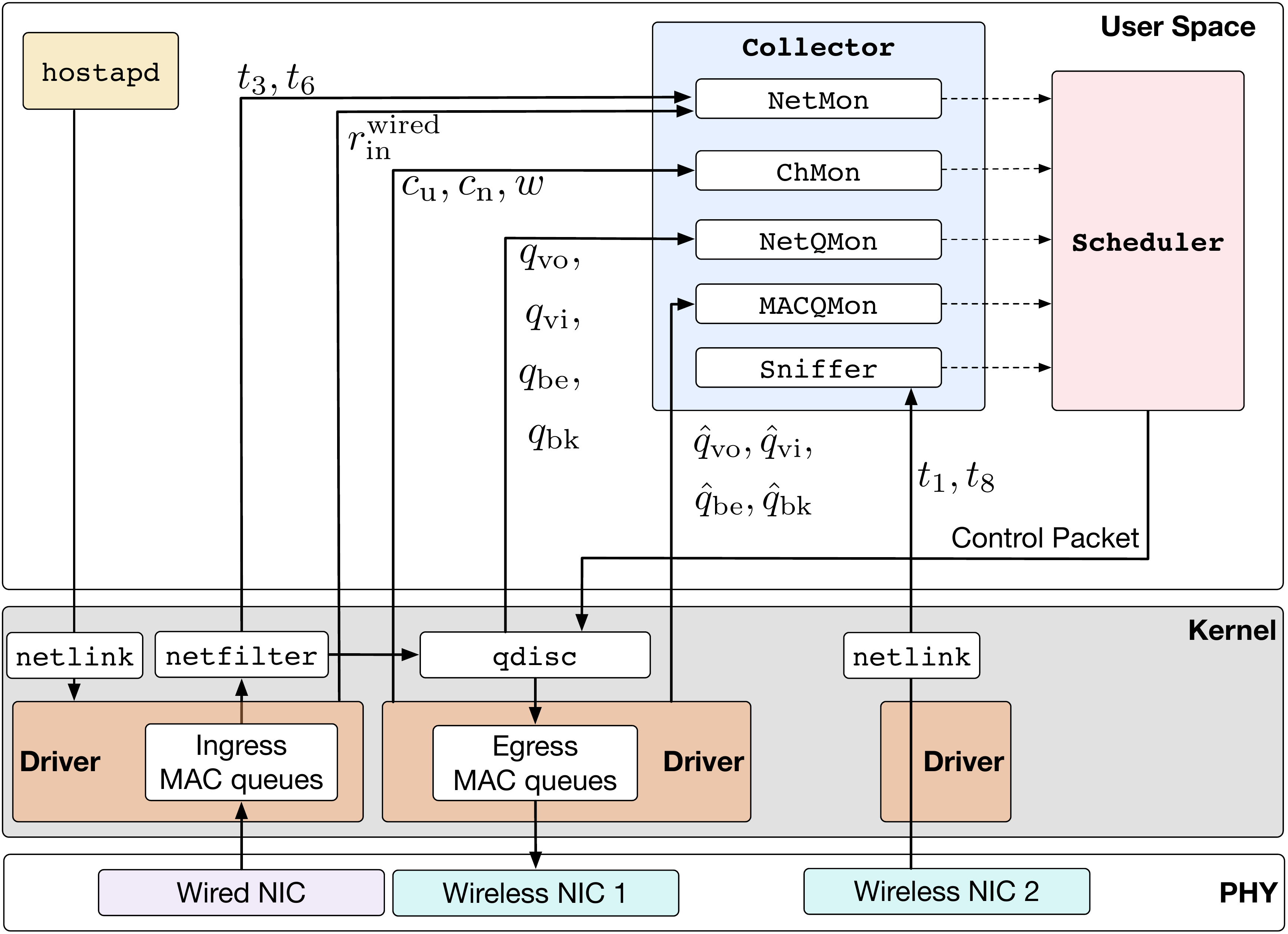}
  \caption{The \gls{AP} architecture developed and used in this paper. The \Collector \ module communicates with various kernel and user-space components to collect a set of features required for delay prediction. The \Scheduler \ estimates the sleep duration and conveys it to the station.
  This figure primarily focuses on the wired-to-wireless interfaces path to compute $\delta_{c}$. Some of the modules required to collect other delay components ($\delta_{a}$ and $\delta_{b}$) are not included in this figure.
  }
  \label{fig:sys_arch}
\end{figure}

Unfortunately, the current \gls{AP} architectures do not provide the necessary tools to collect and apply predictive scheduling \cite{wilhelmi2020flexible}.
In this section, we present an \gls{AP} architecture that allows us to collect the features necessary for predictive scheduling.
Figure \ref{fig:sys_arch} presents the modules we developed on a Linux-based \gls{AP}.
The user-space components of the AP are \Collector{}  and \Scheduler.
The \Collector{} is responsible for collecting all the features required to predict delay.
This information is then shared with and used by the \Scheduler{} to train a model, estimate the sleep duration, and dispatch the schedule.
The \Collector{} module includes the following modules:
The \Sniffer{} module utilizes the \texttt{libpcap} library to capture the timestamp of packets as soon as they are sent or received by the wireless NIC.
The \NetMon{} module records packet exchange instances over the wired interface as well as incoming data rate over this interface.
The \NetQMon{} and \MACQMon{} are responsible for keeping track of the utilization of qdisc and MAC layer queues, respectively.
The \ChUMon{} module captures channel utilization.


To perform the standard \gls{AP} functionalities, we use \texttt{hostapd} \cite{hostapd}, which is a user-space daemon that handles beacon transmission, authentication, and association of stations. 
The underlying hardware includes an Atheros AR9462 wireless NIC, ath9k driver, a Core i3 processor, and 8 GB of RAM.
The \gls{AP} operates in 802.11n mode, uses two antennas, and supports up to 144 Mbps.

We explain the implementation detail and operation of the \gls{AP} in the next three sections.

\subsection{Communication Delay Between \gls{AP} and Server}
Once the \gls{AP} receives an uplink packet, it is stored in the Linux kernel's \textit{qdisc} buffers, and then is sent over the wired interface.
The qdisc is the scheduling mechanism employed by the kernel to schedule the transmission of packets while switching them between two interfaces.
This buffering delay, denoted as $\delta_{a} = t_{3} - t_{1}$, depends on the difference between the rate of incoming wireless uplink packets (destined to the wired interface) and the rate of transmitting these packets over the wired interface.
In this paper, we assume that the speed of the wired interface is fixed. 
This is a reasonable assumption:
First, in enterprise environments, \gls{AP}s are connected to switches via Ethernet links supporting at least 1 Gbps. 
This may also be true in a residential environment where the \gls{AP} is connected to a local processing server through Ethernet \cite{wei2013rt}.
For residential environments, also, fiber-to-the-home (FTTH) provides data rates higher than wireless.
Second, the uplink speed between a home modem and an Internet provider is fixed.
For example, DOCSIS employs a combination of TDMA and CDMA for deterministic channel access.

Based on these observations, we compute $\delta_{a}$ by knowing the number of packets currently in the qdisc\footnote{qdisc implements a scheduler responsible for traffic shaping and policing at layer 3.} of wired interface (\textit{not shown} in Figure \ref{fig:sys_arch}).
We have modified the qdisc module to communicate the number of packets in this buffer with \NetMon.
For each packet $p_{i}$ in qdisc, the \Scheduler{} computes $\nu(p_{i}) = 8 \times (s(p_{i}) + h_{\mathrm{mac}} + h_{\mathrm{phy}})/l_{\mathrm{out}}^{\mathrm{wired}}$, where $\nu(p_{i})$ is the time required to transmit $p_{i}$, $s(p_{i})$ is the packet size (bytes), $h_{\mathrm{mac}}$ is the MAC header size (bytes), $h_{\mathrm{phy}}$ is the physical header size (bytes), and $l_{\mathrm{out}}^{\mathrm{wired}}$ is the transmission bit rate supported by the wired link.
The switching delay is therefore computed as $\delta_{a} = \sum_{\forall p_{i} \in \mathrm{qdisc}} \nu(p_{i})$.

The delivery delay between the \gls{AP} and the server, i.e., $t_{4} - t_{3}$ and $t_{6} - t_{5}$, depend on various factors and primarily on the number of hops between these two nodes.
Based on this number, we consider \textit{edge} and \textit{cloud} computing scenarios.
Edge computing is employed in latency-sensitive and mission-critical applications to minimize the latency and overhead of communication over the wired network \cite{powell2019fog}.
In the cloud computing scenario, the server is located at least a few hops away from the \gls{AP}.
For both cases, to measure this delay, we use a moving average, which is the standard approach used by various protocols such as TCP to estimate RTT \cite{jacobson1988congestion,ludwig2000eifel}.
To this end, we modify the \texttt{netfilter} \cite{russell2002linux} kernel module to communicate with the \NetMon{} module and timestamp $t_{3}$ as the instance the packet is sent to the wired NIC, and $t_{6}$ as the instance the packet has arrived in the \gls{AP}.




\subsection{\gls{AP} to Station Delivery Delay}
\label{sec-ap-sta-del}
An incoming packet from the wired interface first passes through ingress driver queues. 
Subsequently, the packet is processed by the \netfilter{} module.
The packet is then queued in the qdisc.
Finally, the packets are queued at the \gls{EDCA} queues inside the wireless NIC's driver. 
These packets are served according to the channel contention parameters specified by the 802.11e standard.
Specifically, each of the driver queues contends (individually) for channel access before packet transmission.

%
Here we mainly focus on the delay between the arrival of a downlink packet through the \gls{AP}'s wired interface and its transmission through the wireless interface.
This delay is denoted as $\delta_{c} = t_{7} - t_{6}$.
It is particularly challenging to model and predict this delay because it is affected by several factors such as queuing strategy and queue utilization at the qdisc and MAC layer, channel utilization, number of stations, and link quality. 
However, in addition to the high complexity of buffering mechanisms implemented by wireless drivers such as ath9k and ath10k \cite{hoiland2017ending,showail2016buffer}, the actual operation of non-open source drivers is not known, which makes it impossible to develop a mathematical model of buffering delay.
Therefore, we follow a data-driven approach to predict $\delta_{c}$. 
The predicted value is denoted as $\delta_{c}^{\prime}$.
The \Collector{} module time stamps the switching delay between the wired and wireless interfaces.
The time of packet arrival from the downlink transmission is determined by the \Sniffer{} module, which in turn informs the \Collector.
During $t_{6}$ to $t_{8}$, the \Collector{} also collects statistics regarding the status of queues and channel condition.
The collected parameters are explained in the following sections.


\subsubsection{Input Traffic Rate through Wired Interface}
The incoming traffic through wired interface, denoted as $r_{\mathrm{in}}^{\mathrm{wired}}$ (bytes/second), impacts the current and future utilization of \gls{AP}'s layer-3 and layer-2 queues. 
Especially, the \textit{burstiness} of the traffic cannot be characterized by taking into account only queue utilization in the \gls{AP}s. 
Hence, the \NetMon{} module communicates with wired interface driver to collect incoming traffic rate.


\subsubsection{qdisc Queues}
The Linux's network layer buffering of packets during wired to wireless switching is administered by qdisc.
By default, every network interface is assigned {pfifo\_fast} as its transmission queuing mechanism \cite{hoiland2015good}.
This mechanism contains three bands, and dequeuing from a band only happens when its upper bands are empty.
The {PRIO} qdisc is a classful-configurable alternative of pfifo\_fast and enables us to configure the number of bands.
To enqueue the packets of each \gls{AC} in its own queue, we implement four queues in this layer.
These queues are denoted by ${\mathbf{Q}} = \{{q}_\mathrm{{vo}}, {q}_\mathrm{{vi}}, {q}_\mathrm{{be}}, {q}_\mathrm{{bk}} \}$.
We have modified the PRIO kernel module to communicate with the \NetQMon{} module to collect the number of packets in each qdisc band.

With PRIO qdisc, the queuing delay experienced by a packet enqueued in the lowest priority queue not only depends on the current utilization level of that queue, but also on the number of packets in the higher priority queues. 
In addition to the four queues mentioned above, we have also included an additional queue---called \textit{control queue}---that is assigned the highest priority level.
We will utilize this queue in order to implement a higher priority data plane used to send the \textit{control packet} that conveys sleep schedules to stations.
We will explain this packet later.
It is worth mentioning that, although our implementation includes only the PRIO qdisc mechanism (the default policy used in several Linux systems) the concept can easily be extended to other types of qdisc scheduling strategies as well.


\subsubsection{Wireless Channel Condition}
Both interference and channel utilization are the main channel condition parameters that affect packet transmission delay.
However, the duration and intensity of these parameters depend on various factors, such as the number of contending stations and \gls{AP}s, burst size, TXOP, and the transmission power of nearby stations and \gls{AP}s. 
Therefore, accounting for the effect of channel condition through measuring the factors (mentioned above) would be very challenging.
Instead, we collect three parameters to capture the effect of interference and channel utilization on the delay of packet transmission.
The first parameter is channel utilization (${c_\mathrm{u}}$), which refers to the amount of time the \gls{AP} or its associated stations are transmitting.
The second parameter is the number of MAC layer retransmissions ($w$) performed by the \gls{AP} to deliver packets to stations.
The third parameter is the channel's noise level (${c_\mathrm{n}}$), which reflects the activity of nearby wireless devices (such as other \gls{AP}s, ZigBee, and Bluetooth devices).

Most 802.11 drivers maintain counters that represent traffic and channel conditions.
For example, the rate of updating \texttt{ch\_time\_busy} reflects channel utilization during a sampling interval.
The \ChUMon{} module is responsible to extract these counters from wireless NIC driver.

%
%

However, we realized that the interval of obtaining channel utilization also impacts measurement accuracy.
We obtained the peak accuracy, in terms of Kendall's correlation coefficient, when the frequency of polling $c_{\mathrm{u}}$ is 10 ms. 
Additionally, the granularity of the measurements also decreases as we increase the frequency of polling the channel utilization. 
This is because the counters provided by the NIC are reported in milliseconds. 
For example, if the sampling frequency is 10 ms, the granularity of $c_{\mathrm{u}}$ obtained in percentage is 10\%.

\subsubsection{Driver's Transmission Queues}
Using \gls{EDCF}, packets arriving at the MAC layer are categorized and inserted into one of the four queues assigned to each station inside the driver.
The categorization relies on the IP header's \gls{ToS} field.
These queues are denoted by $\hat{\mathbf{Q}} = \{\hat{q}_\mathrm{{vo}}, \hat{q}_\mathrm{{vi}}, \hat{q}_\mathrm{{be}}, \hat{q}_\mathrm{{bk}} \}$.
Each of these queues behaves like a virtual station that contends for channel access independently according to the contention parameters specified in beacon frames.
In case of internal collision between two or more queues, the higher priority queue is granted the transmission opportunity. 
The status of these queues are monitored by the \MACQMon{} module through communicating with the driver.



\subsubsection{Summary of the Features Collected}
The \Scheduler{} interacts with \Collector{} to gather the features necessary for delay prediction.
In summary, the developed \gls{AP} enables us to collect the following features periodically:

\begin{equation}
\label{dataset_entry}
\begin{aligned}
\mathbf{C_{u}}, \mathbf{C_{n}}, \mathbf{R}_{\mathrm{in}}^{\mathrm{wired}}, \mathbf{W},
    \mathbf{{Q}}_\mathrm{{vo}}, \mathbf{{Q}}_\mathrm{{vi}}, \mathbf{{Q}}_\mathrm{{be}}, \mathbf{{Q}}_\mathrm{{bk}},  \\ 
    \widehat{\mathbf{Q}}_\mathrm{{vo}}, \widehat{\mathbf{Q}}_\mathrm{{vi}}, \widehat{\mathbf{Q}}_\mathrm{{be}}, \widehat{\mathbf{Q}}_\mathrm{{bk}}
\end{aligned}
\end{equation}
where 
$\mathbf{C_{\mathrm{u}}}$, 
$\mathbf{C}_{n}$,
$\mathbf{R}$,
$\mathbf{Q}$, 
$\widehat{\mathbf{Q}}$, 
and $\mathbf{W}$
represent the
list of channel utilization values, 
list of channel noise values, 
list of incoming traffic rate values over wired interface,
list of MAC layer downlink retransmission values,
list of qdisc-queue utilization values (for each \gls{AC}),
and list of driver-queue utilization values (for each \gls{AC}), respectively.
Each of these lists includes the values that have been periodically collected.
For example, assuming that each list contains $k$ values, list $\mathbf{C}_{u}$ is represented as follows:
\begin{equation}
\label{expanded_ataset_entry}
\begin{aligned}
\mathbf{C}_{\mathrm{u}} = \lbrack {c_\mathrm{u}}({t^{\prime} - ({k-1}) \times \Delta}), \\
{c_\mathrm{u}}({t^{\prime}-({k-2})\times \Delta}),..., 
{c_\mathrm{u}}({t^{\prime}-\Delta}),
{c_\mathrm{u}}(t^{\prime}) \rbrack
\end{aligned}
\end{equation}
where ${c_\mathrm{u}}(t^{\prime})$ is the last sampled value based on the periodical sampling technique, and $\Delta$ refers to sampling interval.
In our implementation, we collect samples every 10 ms ($\Delta = 10$ ms), and each prediction uses the past 5 sampled values ($k = 6$).
We did not use a shorter sampling interval because that resulted in a high processor utilization ($> 30\%$).
Implementing a more efficient \gls{AP} architecture is one of our future works.


In addition to the features collected periodically, we add two features that are computed once for each prediction.
First, since each AC is treated differently at the driver, we include the AC of the transaction as a feature as well.
The second feature is $\delta_{a}$ + $\delta_{b}$. 
Including this feature is motivated by the fact that the predicted delay ($\delta_{c}'$) depends on the 
interval between the uplink packet and the arrival of downlink packet over the wired interface.
For example, if the server delay is expected to be 30 ms, the prediction for $\delta_{c}$ should be made for a packet that would arrive at the AP after 30 ms.

\subsection{Schedule Announcement}
\label{scheduledistribution}

The \Scheduler{} module is also responsible for conveying the predicted sleep schedule to the station.
Once the prediction has been made, the \Scheduler{} creates a UDP control packet and sends it to the station.
This packet includes $\delta_{a}$, $\delta_{b}$, and $\delta_{c}$, as well as the standard deviation of prediction error, where each is encoded as one byte. 
The value of each byte reflects duration in milliseconds.
To prioritize the transmission of this tiny packet over existing data packets, we modify the qdisc module and add a high-priority band to it.
The \gls{ToS} field of this high-priority band is 7, which is the highest network layer priority.
This band is also mapped to the voice \gls{AC} of driver's queues to expedite packet transmission.
This \gls{AC} uses a shorter contention window and smaller \gls{AIFS}, as defined by the 802.11e standard.
Therefore, all the control packets generated by the \Scheduler{} are sent through this high-priority data plane.
Once this control packet reaches the station at $t_{2}$, the station immediately transitions into sleep state for $\delta_{a} + \delta_{b} + \delta_{c} - (t_{2} - t_{1})$.
Note that the station simply uses a timer to measure $t_{2} - t_{1}$.

To further reduce the \gls{AP}'s processing overhead and the idle listening time of station while waiting for the control packet from the \gls{AP}, we use \texttt{mmap} \cite{mmap} for fast and shared access.
In other words, the most recent information collected by the \Collector{} (which are shared with the \Scheduler) is stored in the physical memory to eliminate the unpredictability of accessing this data through files.

At the end of the sleep interval, the station wakes up and informs the \gls{AP} about its transition into the awake mode. 
This is achieved by relying on \gls{APSD}, which is supported by the state-of-the-art wireless NICs.
To this end, at $t_{7}$, the station wakes up and sends a NULL packet to the \gls{AP}, conveying that the station is ready for receiving a packet. 
The \gls{AP} responds by sending the downlink packet (or starting a \gls{TXOP}) at $t_{8}$.

\section{Predictive Scheduling}
\label{Predictive Scheduling}
In this section, we first present our testbed setup for realistic traffic generation necessary for training and evaluation.
Utilizing this dataset, we discuss various stages of the statistical learning and modeling process. 
Finally, we compare the performance of several machine learning approaches for delay prediction.

\subsection{Traffic Generation}
\label{datacollection}
As explained in the previous sections, \gls{AP} modification is necessary to collect the features required for predictive scheduling.
Also, we need to introduce controlled changes in the traffic pattern of the environment to study the impact of these changes on prediction accuracy.
Therefore, it is required to have a testbed that represents the traffic pattern of real-world environments as well as controllability over traffic generation parameters.
To achieve this, we systematically characterize and compare the scenarios generated in our testbed with those collected in real-world environments.

\begin{figure}
\centering
  \includegraphics[width=0.98\linewidth]{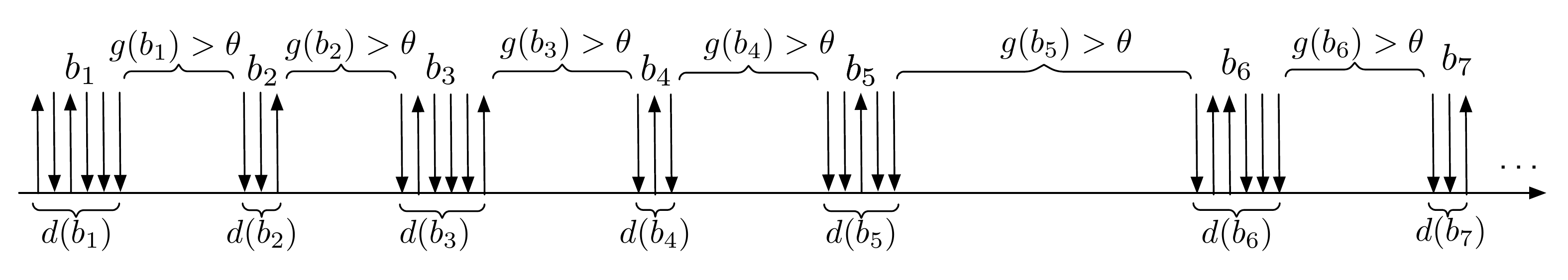}
  \caption{Traffic characterisation. }
  \label{fig:bursts}
\end{figure}

A \textit{burst}, denoted as $b_{i}$, is defined as a train of packets in either UL or DL direction with inter-arrival time less than a threshold value $\theta$ \cite{lan2006measurement}.
Resembling 802.11 traffic, Figure \ref{fig:bursts} illustrates a series of bursts. 
The duration (in seconds) of a burst $b_{i}$ is denoted by $d(b_{i})$. 
And $g(b_{i})$ refers to the gap (in seconds) between two consecutive bursts $b_{i}$ and $b_{i+1}$.

In order to generate traffic that represents various levels of \textit{network dynamics} in real-world environments, we have developed a testbed that includes two categories of stations: 
(i) stations such as laptops, phones, and IoT devices, and 
(ii) four Raspberry Pi boards to control traffic generation pattern.
Each RPi runs four threads, where each thread can be involved in a downlink, uplink, or bidirectional flow.
This enables us to introduce up to 16 additional controlled flows into the network.
The implementation of traffic control capability is composed of a set of Python scripts that use the iperf tool under the hood.
A central controller is in charge of setting the parameters of traffic flows.
Among the flow parameters, we can modify the access category, transport layer protocol, packet size, bit rate, burst size, and inter-burst interval.
Also, we note that sharing flow characteristics by consecutive flows is more likely.
To represent this behavior, after each burst, the controller either repeats the process of traffic selection or chooses the same parameters for the next burst based on a \textit{variability parameter} denoted by $\nu$.
Specifically, a higher value of $\nu$ results in a higher dynamicity.
Hence, we use $\nu = 0.9$ for generating flows that resembles \textit{high dynamicity} (HD) traffic and $\nu = 0.1$ for generating less diverse traffic referred to as \textit{normal dynamicity} (ND).
Also, for the voice and video \gls{AC}s, UDP is preferred because it is the dominant transport protocol for real-time traffic.

As demonstrated in \cite{lan2006measurement}, capturing network dynamics can be achieved by focusing on characterizing burstiness.
Additionally, Xiao et al. \cite{xiao2013modeling} characterized a flow as \textit{regularly} bursty when the standard deviation of the inter-burst intervals ($g(\cdot)$ seconds) and burst sizes ($s(\cdot)$ bytes) are relatively smaller. 
Otherwise, the flow is characterized as \textit{randomly} bursty.
However, based on Xiao's metrics for burstiness, traffic with fewer bursts per unit time can still have a high standard deviation of $s(\cdot)$ and $g(\cdot)$.
Hence, we consider burst frequency (i.e., number of bursts per second) and burst size for calculating traffic burstiness.
Additionally, due to the difference in the scale of those two parameters, we normalize the burst rate (per second) in the range $\left[0, 1\right]$. 
We define traffic \textit{burstiness}, denoted by $\mathcal{B}$, as follows:
\begin{equation}
\label{dataset_entry}
\begin{aligned}
    \mathcal{B} = \left(1 - \frac{1}{\mathcal{M}}\right) \times \left(\frac{\sum_{ i=1}^{N} s(b_i)}{N}\right) \\
\end{aligned}
\end{equation}
where $N$ is the number of bursts in the dataset, $s(b_{i})$ is the size of burst $b_{i}$ (in bytes), and $\mathcal{M}$ is average number of bursts per second.
%
%

In addition to traffic burstiness, we define another metric that represents traffic \textit{dynamicity} based on various aspects including burst size, burst duration, inter-burst interval, and the access category of the packets in each burst.
This metric, which we refer to as \textit{dynamicity} and denoted by $\mathcal{D}$, is defined as follows:

\begin{equation}
\label{dataset_entry}
\begin{aligned}
\\
\mathcal{D} = \left({\frac{1}{N}}\times \sum_{i = 2 }^{N}\frac{\frac{|d(b_i) - d(b_{i-1})|}{d(b_{i-1})}}{g(b_i)}\right) + 
\\
 \left({\frac{1}{N}}\times \sum_{i = 2 }^{N}\frac{\frac{|s(b_i) - s(b_{i-1})|}{s(b_{i-1})}}{g(b_i)}\right) + 
\\
\left({\frac{1}{N}}\times \sum_{i = 2 }^{N}\frac{\frac{|p(b_i) - p(b_{i-1})|}{p(b_{i-1})}}{g(b_i)}\right) +
\left({\frac{1}{N}}\times \sum_{i = 2 }^{N}\frac{z(b_i)}{g(b_i)}\right) 
\end{aligned}
\end{equation}
where,
\begin{equation}
\label{dataset_entry}
\begin{aligned}
z(b_i) =  
{\frac{|p_\mathrm{vo}(b_i) - {p}_\mathrm{vo}(b_{i-1})|}{{p}_\mathrm{vo}(b_{i-1})}} +
{\frac{|{p}_\mathrm{vi}(b_i) - {p}_\mathrm{vi}(b_{i-1})|}{{p}_\mathrm{vi}(b_{i-1})}} +
\\
{\frac{|{p}_\mathrm{be}(b_i) - {p}_\mathrm{be}(b_{i-1})|}{{p}_\mathrm{be}(b_{i-1})}} +
{\frac{|{p}_\mathrm{bk}(b_i) - {p}_\mathrm{bk}(b_{i-1})|}{{p}_\mathrm{bk}(b_{i-1})}}
\end{aligned}
\end{equation}
Here, $p_{x}(b_i)$ is number of packets belonging to an \gls{AC} $x$ in a burst $b_{i}$. 
Parameter $z(b_{i})$ reflects the change in the number of packets belonging to each access category in each burst compared to that in the previous burst.


\subsection{Data Collection}
\begin{figure}
\centering
   \includegraphics[width=0.98\linewidth]{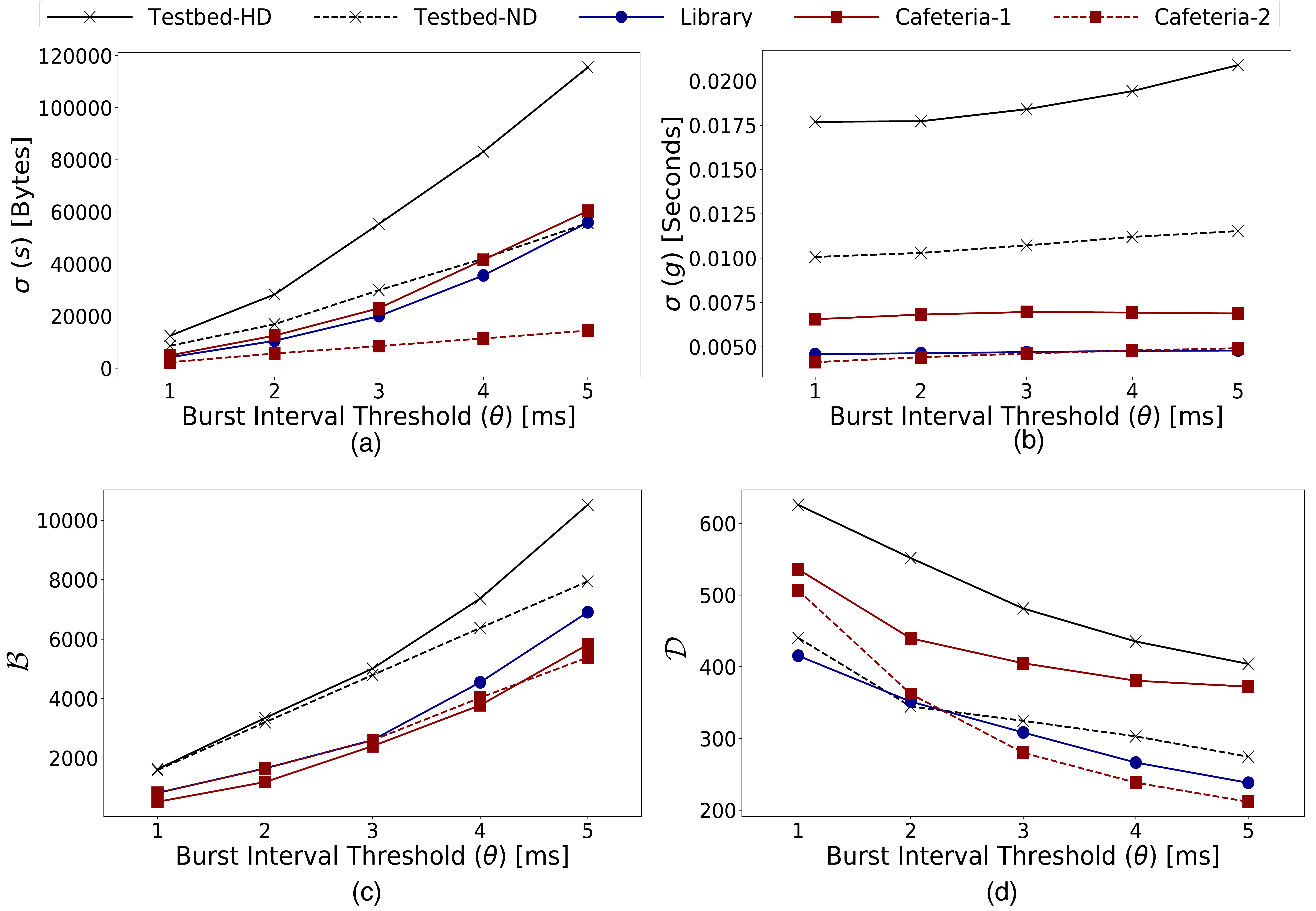}
  \caption{(a) Standard deviation of burst size, (b) standard deviation of burst interval, (c) burstiness ($\mathcal{B}$), and (d) dynamicity ($\mathcal{D}$) of traffic generated by our testbed compared to traffic captured in real-world environments. ND and HD refer to normal and high dynamicity, respectively.
  }
  \label{fig:burstiness_and_dynamicity_graphs}
\end{figure}

Using the metrics mentioned above, we compare our testbed generated datasets against those collected in multiple real-world environments.
Figure \ref{fig:burstiness_and_dynamicity_graphs} presents the results.
In general, we can observe that our ND scenario resembles real traffic. 
The HD scenario offers higher network dynamics, which is essential to study the robustness of predictive scheduling.

When generating data in our testbed, the type of each transaction is selected from the voice, video, background, and best-effort \gls{AC}s with equal probability.
The inter-transaction delays are uniformly distributed between 1 ms and 500 ms. 
In addition to the features discussed in Section \ref{delay_analysis}, we also collect $\delta_{a}$, $\delta_{b}$, and $\delta_{c}$ values per transaction.
We split each dataset, such that 70\% of it is used for training and the remaining 30\% is used for validation.
We use independent datasets consisting of 10,000 data points for evaluating the performance and robustness of each modelling approach (test datasets) in the ND and HD scenarios.


%

\subsection{Data Pre-processing}

We focus on delay prediction for $\delta_{c} < 100$ ms, for two reasons:
First, considering edge computing scenarios, observing RTTs more than 100 ms is very unlikely.
Second, almost all commercial \gls{AP}s implement 102.4 ms as their beaconing period.
Therefore, all stations wake up every 102.4 ms to synchronize with \gls{AP} beacons and check if the \gls{AP} has any buffered packets.


Although we have ascertained that there is no unnecessary user-space applications running on the \gls{AP}, running kernel tasks, as well as context switching of existing processes, can lead to a higher operating system response time.
Hence, we have noticed that the latency of data collection from the software modules discussed in Section \ref{delay_analysis} is impacted occasionally and results in longer polling intervals.
We filter the datasets to remove such \textit{stale} feature values that do not reflect the latest status of the channel or the queues.


The feature set varies in terms of ranges and units.
For example, $c_{\mathrm{u}}$ varies from 0 to 100\%, whereas $c_{\mathrm{n}}$ varies from $-95$ dBm to $-66$ dBm.
Since this would result in disproportional treatment of the features by the machine-learning algorithms, we scale each of the features into the range $\pm1$.
Furthermore, the dataset contained more samples (transactions) whose actual delay is within the range $[1, 50]$ ms, compared to the samples whose actual delay was between $[50, 100]$ ms. 
We under-sample the majority bins to prevent the machine-learning algorithm from generalizing the results for the packets whose actual delay is higher.

\subsection{Regression models}
Given the continuous nature of the target variable, we identify the problem of predicting $\delta_c$ as a regression problem. 
The methods that we considered are \gls{RFR}, \gls{GBR}, \gls{ETR} and \gls{HBR}, which are widely-used ensemble learning methods for regression.
We also considered (deep) neural networks, which have been shown to be effective in different areas of machine learning such as prediction in time-series data and image processing. 

In ensemble learning, final prediction can either be calculated by average of the predictions of the model trained on random subsets of data (bagging) or calculated via sequentially training the model using prediction success on the previous sample of the dataset (boosting).
\gls{RFR} is an example of bagging approach and functions by constructing several decision trees during training and makes predictions based on the outputs of the individual trees. 
\gls{RFR} is a popular algorithm and runs efficiently on large and high-dimensional datasets. 

%
\gls{GBR} is an example of boosting approach.
Each tree outputs a prediction value at different splits that can be added together, allowing subsequent models to modify error in predictions. 
\gls{HBR} is a variant of \gls{GBR}. 
Since it is a histogram-based estimator, HBR can reduce the number of splitting points by binning input samples, and therefore improves the performance when dealing with large datasets. 
\gls{ETR} creates decision stumps at variable tree depths. 
The features and splits are selected randomly, and are less computationally expensive than other tree-based algorithms.

%
%

\gls{NN} have been studied extensively in the past decade for their efficiency in learning complex data features for making predictions.
\gls{RNN} are a class of neural networks in which the predictions are based on the current and also past inputs, and therefore they are suitable for making predictions about $\delta_c$ using historical network features. 
A specific variant of \gls{RNN} is \gls{LSTM} \cite{hochreiter1997long}, which is able to track long-term dependencies of output predictions on input history. 
We use \gls{LSTM} for predicting $\delta_c$ and compare its accuracy with \gls{RFR}, \gls{ETR}, \gls{GBR}, and \gls{HBR}.

For the ensemble learning methods, we use scikit-learn library \cite{pedregosa2011scikit} and tune the hyper parameters using grid search and a validation dataset to obtain the highest performance on the training data while not over-fitting. For the \gls{LSTM} model, we use Tensorflow and Keras library \cite{chollet2015keras}. 
Also, we utilize early stopping mechanism (on the validation dataset) to prevent over-fitting. 
The optimal \gls{LSTM} model contains one \gls{LSTM} layer followed by 3 dense layers, each consisting of 20 neurons and \textit{ReLu} activation function. The model was trained using Adam optimiser \cite{kingma2014adam} with learning rate of 0.01.

\subsection{Model Evaluation}
\begin{figure}
\centering
  \includegraphics[width=1\linewidth]{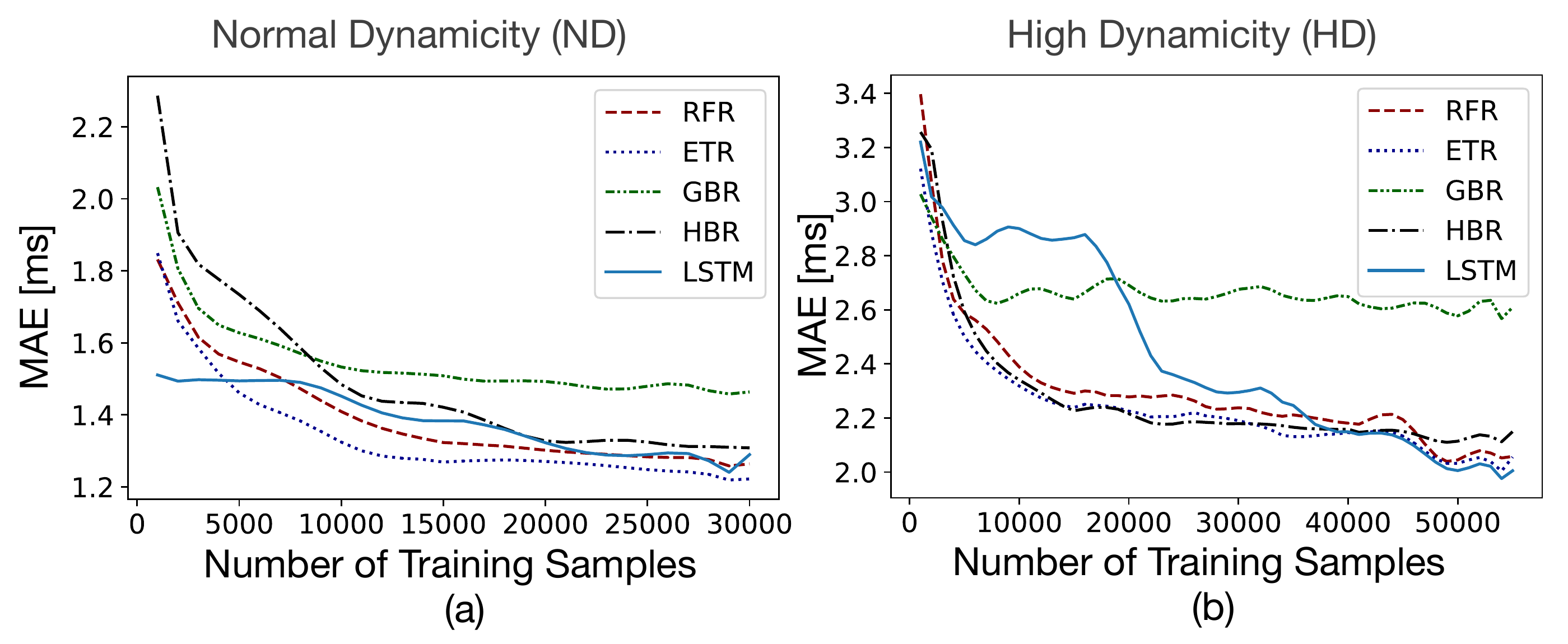}
  \caption{\gls{MAE} of machine-learning algorithms versus the number of samples (transactions) in training dataset for (a) Normal Dynamicity (ND), and (b) High Dynamicity (HD) scenarios. Results are averaged over all \gls{AC}s.
  \gls{ETR} converges the fastest, and \gls{LSTM} requires up to 3x more datapoints.
  }
  \label{fig:lengthandmaer}
\end{figure}

All the evaluations are based on performance of models on the test dataset.
We first investigate the effect of training data size on model's performance.
Figure \ref{fig:lengthandmaer} illustrates the \gls{MAE} of $\delta_{c}$ as a function of the size of training data.
We observed that the performance of \gls{ETR} converges at the fastest rate, utilizing 15000 and 20000 data points for training under the ND and HD traffic, respectively.
Compared to \gls{ETR}, \gls{LSTM} requires as much as 3x more training data for its performance to converge, utilizing the training dataset consisting of 30000 and 50000 datapoints under ND and HD scenarios, respectively.
Therefore, for further evaluation of \gls{LSTM}, we use training datasets consisting of 30000 datapoints for ND scenarios and 50000 datapoints for HD scenarios. 
For other algorithms, we use training datasets consisting of 15000 and 20000 datapoints for ND and HD scenarios, respectively.

\begin{figure}
\centering
  \includegraphics[width=1\linewidth]{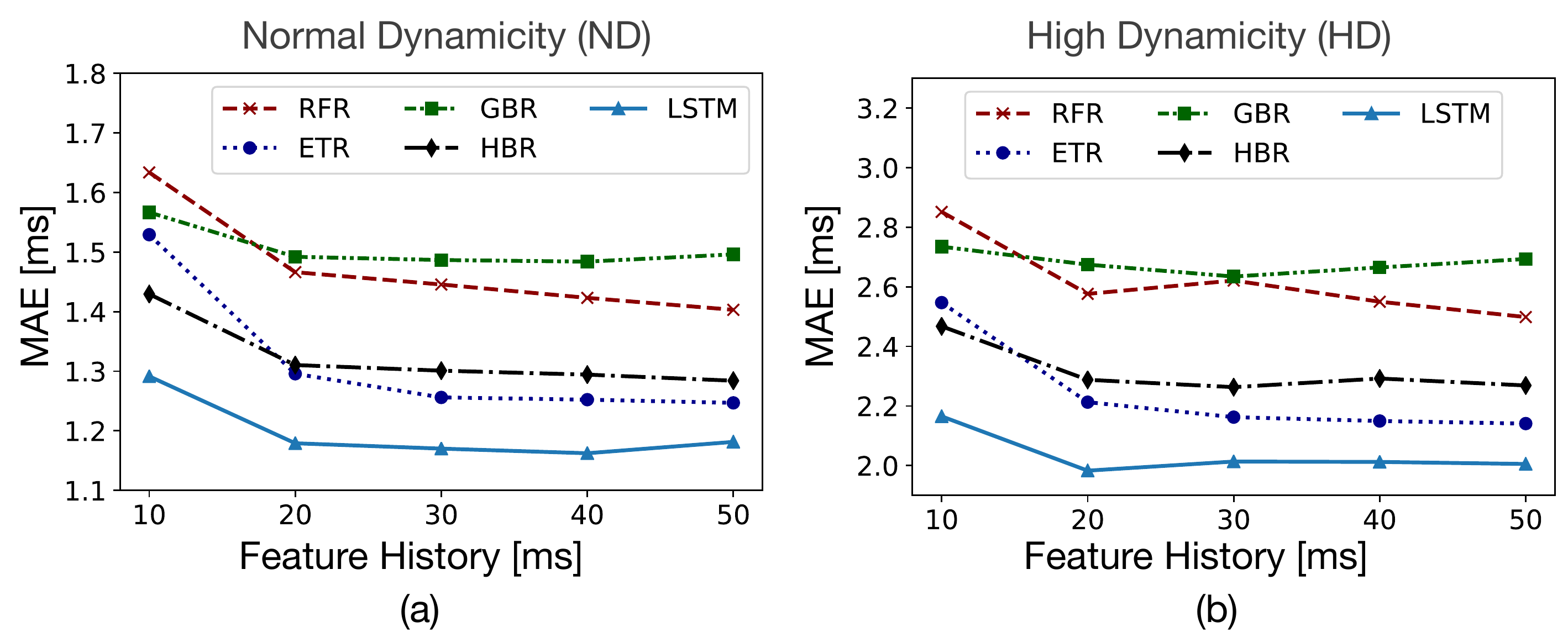}
  \caption{\gls{MAE} of machine-learning algorithms with respect to feature history in (a) Normal Dynamicity (ND), and (b) High Dynamicity (HD) scenarios. Results are averaged over all \gls{AC}s.}
  \label{fig:featurehistory}
\end{figure}

Next, we quantify the effect of feature-history (represented by $k$ in Eq. \ref{expanded_ataset_entry}) on the overall performance of all algorithms. 
Figure \ref{fig:featurehistory} shows the results.
\gls{MAE} decreases significantly in both HD and ND scenarios for all the algorithms when we include the recent feature history as an input for training the model.
This decrease continues up to 30 ms, beyond which it does not result in performance enhancement.
Historical data of features helps the model to anticipate trends of features and accurately predict $\delta_{c}$ that would be incurred by the downlink packet shortly. 
Therefore, in addition to the most recent record, we include the three preceding feature values (corresponding to a total of the preceding 40 ms of feature values) to train the models.

\begin{figure}
\centering
  \includegraphics[width=1\linewidth]{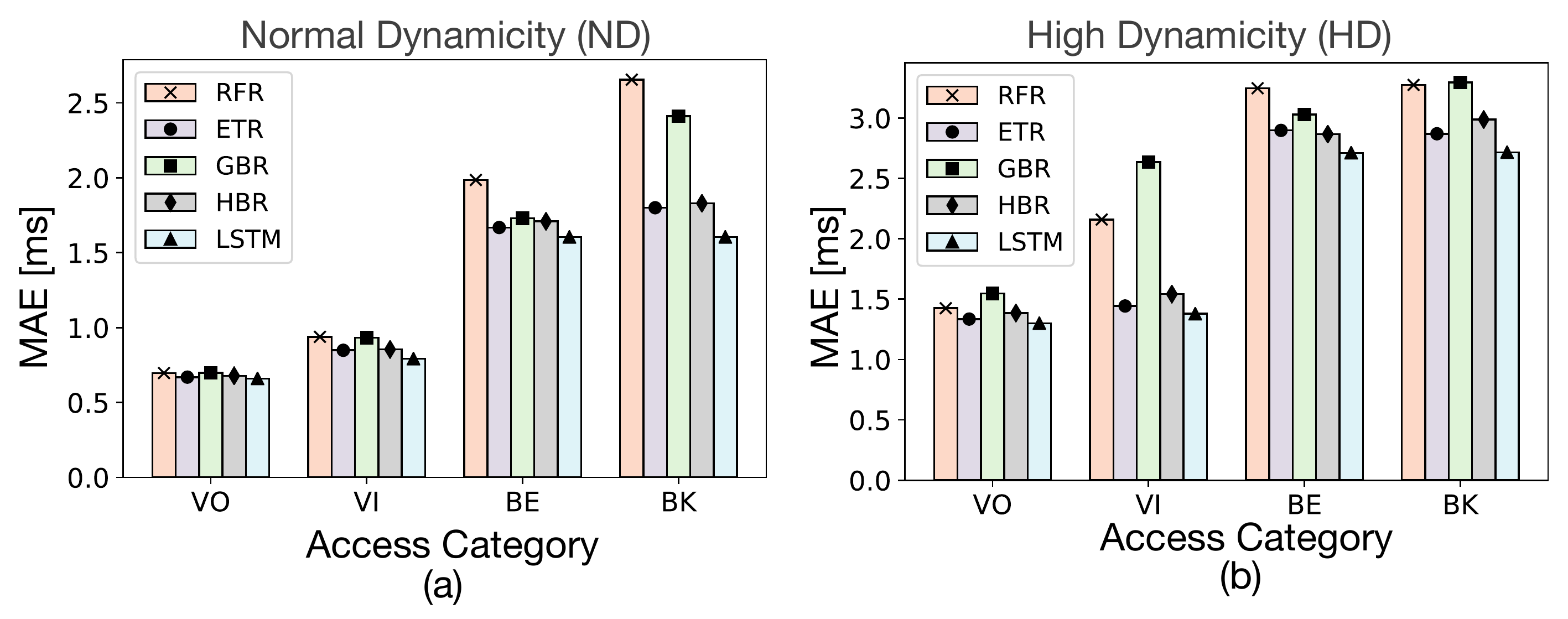}
  \caption{\gls{MAE} of machine-learning algorithms versus transaction's AC in (a) Normal Dynamicity (ND), and (b) High Dynamicity (HD) scenarios.
  \gls{MAE} of VO and VI packets is lower than BK and BE packets.
  }
  \label{fig:ACcompare}
\end{figure}

Figure \ref{fig:ACcompare} compares the performances of the machine-learning algorithms versus \gls{AC} of transactions.
%
In the ND scenario, compared to the average \gls{MAE} of all the other machine-learning algorithms, the \gls{MAE} of \gls{LSTM} is 17\% lower for all ACs.
Similarly, under HD traffic, the \gls{MAE} of \gls{LSTM} is 13\% lower than the average \gls{MAE} of all the other machine-learning algorithms.
This figure also shows that the \gls{MAE} of VO and VI packets is lower than BK and BE packets. 
The reason is that the packets of these \gls{AC}s are prioritized over higher \gls{AC}s at the qdisc layer (using PRIO qdisc) as well as the driver's queues (using \gls{EDCA}).
This results in lower delays incurred by the downlink packets.

Figure \ref{fig:ACcompare} also shows the effect of dynamicity on MAE.
Averaged over all \gls{AC}s, the \gls{MAE} of \gls{RFR}, \gls{ETR}, \gls{GBR}, HBR, and \gls{LSTM} is 1.43 ms, 1.26 ms, 1.49 ms, 1.28 ms, and 1.16 ms in ND scenario.
Whereas, \gls{MAE} of \gls{RFR}, \gls{ETR}, \gls{GBR}, HBR, and \gls{LSTM} is 2.49 ms, 2.17 ms, 2.69 ms, 2.27 ms and 2.01 ms in HD scenario.
Since the performance of \gls{LSTM} is, on an average, 16\% better than other algorithms, we use this algorithm for empirical evaluations in Section \ref{perf_eval}.

\begin{figure}
\centering
  \includegraphics[width=1\linewidth]{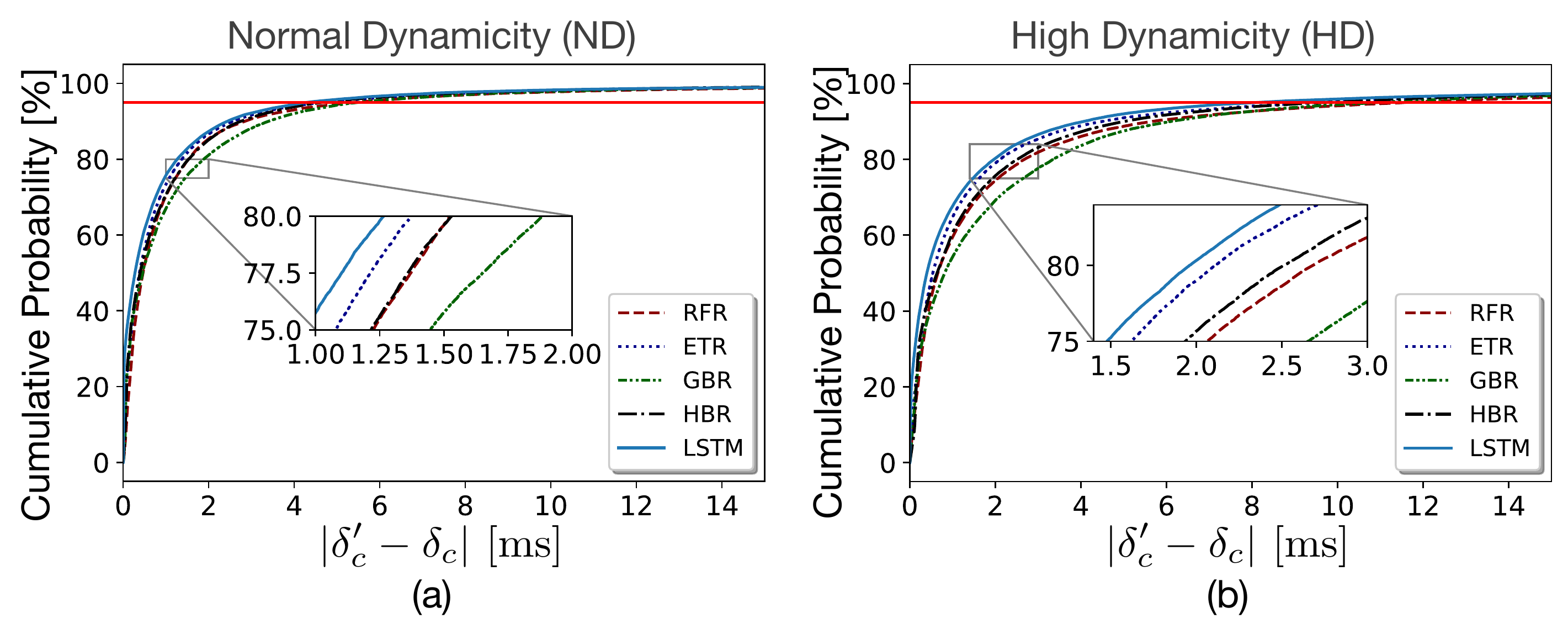}
  \caption{ECDF of prediction errors ($\left|\delta^{\prime}_{c} - \delta_{c}\right|$)  while utilizing various machine learning algorithms in (a) Normal Dynamicity (ND), and (b) High Dynamicity (HD) scenarios. All machine-learning algorithms are able to predict $\delta^{\prime}_{c}$ for 95\% of the packets with an error of $\pm$4.3 ms in case of ND scenario, and $\pm$8.2 ms for the HD scenario.}
  \label{fig:algo_compare}
\end{figure}

Figure \ref{fig:algo_compare} shows the \gls{ECDF} of the deviation of $\delta^{\prime}_{c}$ from $\delta_{c}$.
All machine-learning algorithms are able to predict $\delta^{\prime}_{c}$ for 95\% of the packets with an error of $\pm$4.3 ms in case of ND scenario, and $\pm$8.2 ms for the HD scenario.
However, note that \gls{LSTM} requires about 3x more training data for its performance to converge compared to other algorithms (cf. Figure \ref{fig:lengthandmaer}).
Hence, in scenarios where it is not possible to collect large datasets for training, either \gls{ETR} or \gls{RFR} can also be used.

\begin{figure}
\centering
\begin{minipage}[b]{.45\linewidth}
  \centering
  \includegraphics[width=1\linewidth]{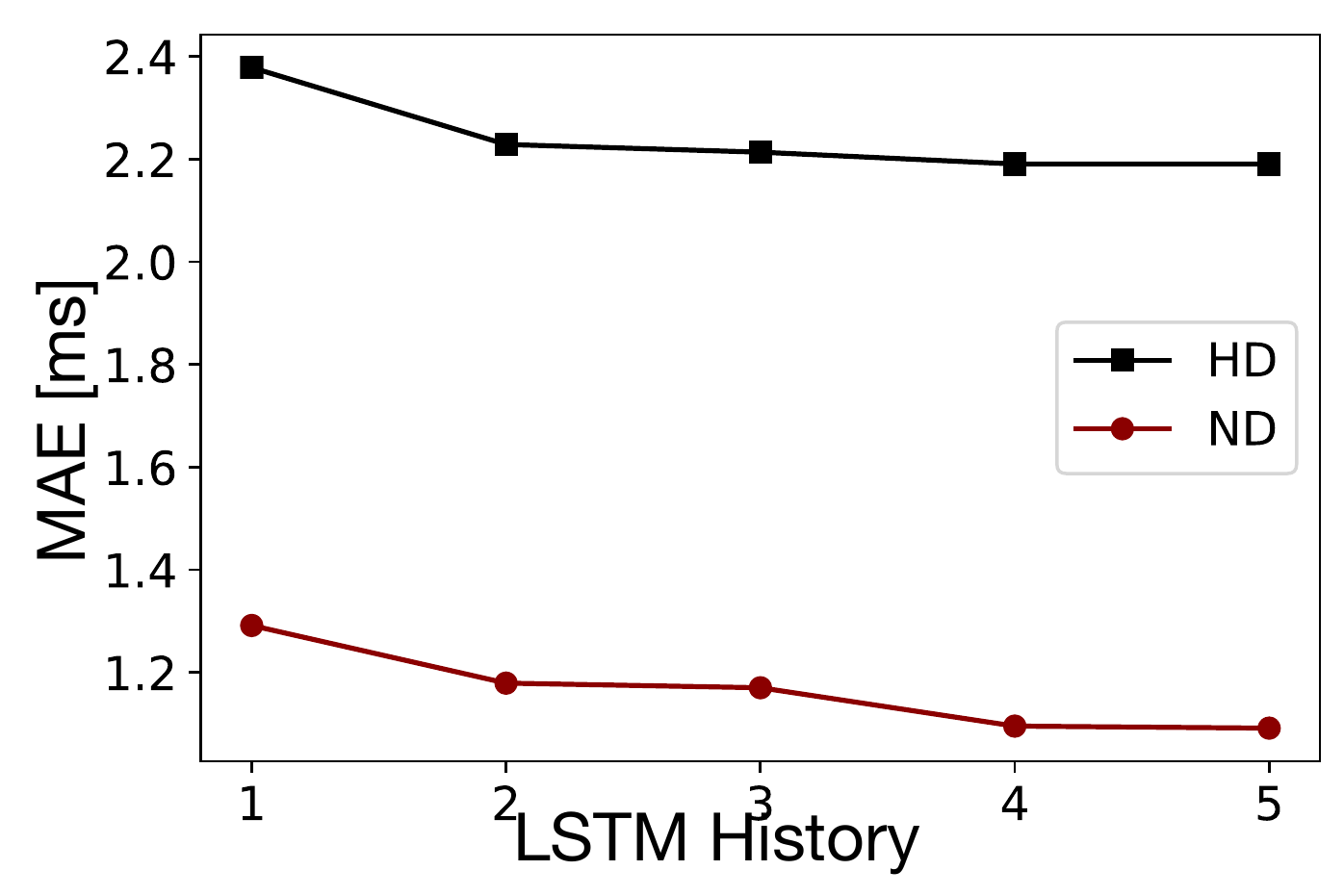}
\caption{\gls{MAE} with respect to \gls{LSTM} history in ND and HD scenarios.}
  \label{fig:LSTMhistory_timeML_single}
\end{minipage}%
\qquad
\begin{minipage}[b]{.45\linewidth}
  \centering
  \includegraphics[width=1\linewidth]{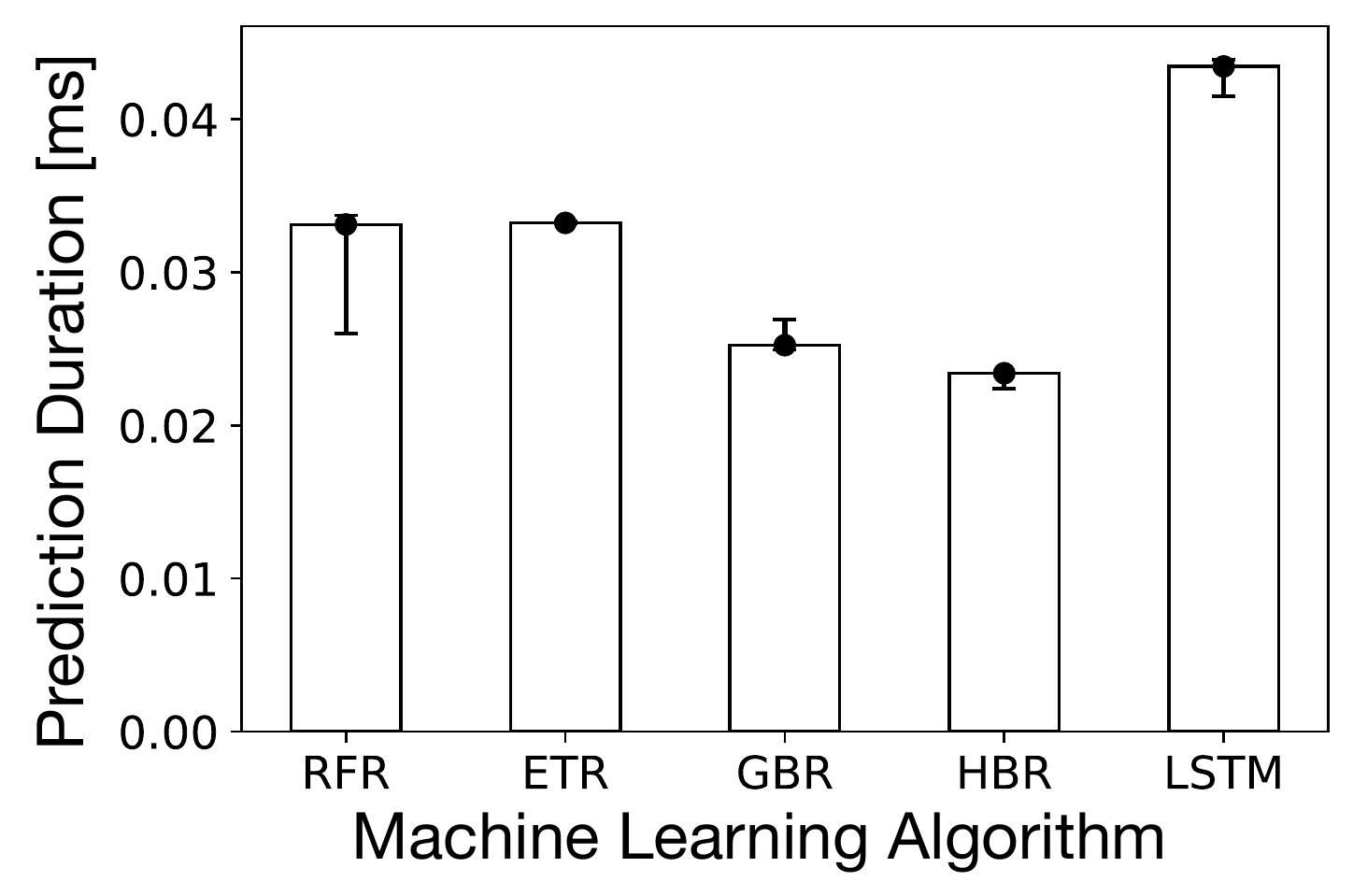}
\caption{Processing time required to predict delay for each transaction.}
  \label{fig:proc_time}
\end{minipage}
\end{figure}

In LSTM networks, neuron activations are dependent on the previous network states. 
The network retains a memory equal to the number of lookbacks (a.k.a., timesteps)---allowing the flow of information from all the previous timesteps \cite{koudjonou2019stateless}.
Lookback is defined as the number of timesteps for which the \gls{LSTM} is unfolded for back-propagation.
Simply put, the number of timesteps reflects the number of previous transactions in the temporal domain that aids in predicting the delay of the current transactions by providing contextual information.
This is particularly beneficial when multiple transactions occur during a single burst or bursts with similar characteristics.
To this end, we estimate the effect of historical data of the previous transactions by including feature values of the past transactions and the current transaction.
Figure \ref{fig:LSTMhistory_timeML_single} shows the \gls{MAE} of \gls{LSTM} versus the number of lookbacks.
We observe that the \gls{MAE} of the \gls{LSTM} model decreases up to five lookbacks. 
This means, on average, five transactions occur during similar network conditions.

%
%

Figure \ref{fig:proc_time} shows the prediction processing overhead of all the machine learning algorithms on a 2.4 GHz Core-i3 processor.
Each marker shows the median of time taken to predict each data point in the test dataset, and the error bars present lower and upper quartiles.
We observed that the HBR algorithm takes the least amount of time (24 $\mu$s median and 0.046 $\mu$s standard deviation) for prediction, whereas \gls{LSTM} requires the most (48 $\mu$s median and 3 $\mu$s standard deviation).
However, the time taken to predict the delay in case of \gls{LSTM} is still considerably shorter than a packet transmission time. 
For example, with a 1400 bytes packet sent over a 54 Mbps link, the ratio of prediction duration to transmission duration is $48 \mu\mathrm{s}/207 \mu\mathrm{s}$.


\textbf{Discussion.}
In our current implementation, we perform both training and modeling on the \gls{AP}. 
However, if the \gls{AP} is not powerful enough to train the model, then the training could be offloaded to a cloud or fog computing platform.
In any case, edge computing is essential to perform scheduling immediately and convey the sleep schedule to the station.



\section{Empirical Evaluation}
\label{perf_eval}

\label{scheduleestimation}
In this section, we present an empirical evaluation of EAPS versus the power saving mechanisms of 802.11.

\subsection{Testbed}
Our testbed includes four IoT stations, four Raspberry Pi boards, regular stations (smartphones and laptops), an \gls{AP}, and a server.
We simply refer to the IoT stations as \textit{station}.
Each station is a Cypress CYW43907 \cite{cypressb}, which is a low-power 802.11n SoC designed for IoT applications.
To represent a real-world scenario affected by variable background interference, the testbed is located in a residential environment surrounded by several \gls{AP}s belonging to different households.
Also, the four Raspberry Pi boards are used to control network dynamicity and variability in $\delta_{c}$.

To represent the request-response behavior of IoT traffic, for each uplink UDP packet sent, the server responds by sending a downlink packet back to the station\footnote{Note that the case where multiple uplink and downlink packets are exchanged is simply supported as explained in Section \ref{delay_analysis}.}.
The exchange of an uplink packet and receiving its response is referred to as a \textit{transaction}. 
In all the figures of this section, each marker shows the median of 1000 transactions and the error bars present lower and upper quartiles.
We use the EMPIOT tool \cite{dezfouli2018empiot} to measure the energy and delay of each transaction.
This tool samples voltage and current at approximately 500,000 samples per second. 
These samples are then averaged and streamed at 1 Ksps. 
The current and voltage resolution of this platform are 100 $\mu$A and 4 mV, respectively.

We use two scenarios to evaluate the performance of EAPS with respect to varying \gls{AP}-server delays (i.e., $\delta_{b}$ in Figure \ref{fig:delays1}): \textit{edge}, and \textit{cloud} computing.
In the former, the server is directly connected to the \gls{AP}, and in the latter, we use an Amazon AWS server in Oregon.
Note that in both cases the sleep schedules are computed at the edge and by the \gls{AP} the station is associated with.

\subsection{Baselines and EAPS Variations}
The baseline mechanisms used are PSM, APSM, and CAM.
Using PSM, after an uplink packet, the station goes back into sleep mode and wakes up at each beacon instance to check for downlink packet delivery.
With APSM, instead of going back into sleep right after uplink, the station stays in the awake mode for 10 ms.
With CAM, the station always stays in awake mode.
\textit{Note that for CAM, to avoid the impact of inter-transaction time on results, we only measure the delay and energy consumption of transactions (only between the uplink and downlink packets).}

\begin{figure}
\centering
  \includegraphics[width=1\linewidth]{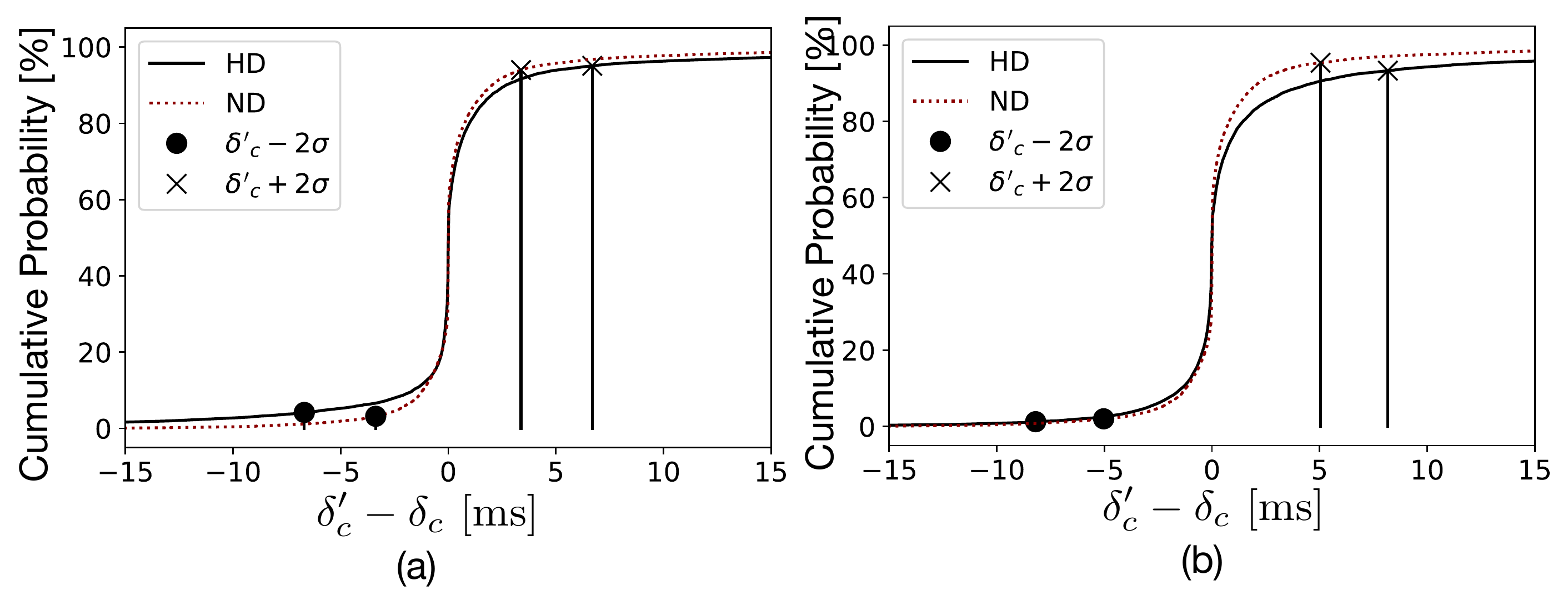}
  \caption{Distribution of prediction error ($\delta^{\prime}_{c} - \delta_{c}$) for (a) voice, and (b) background \gls{AC}s.
  Prediction error of voice \gls{AC} is lower than that of background \gls{AC}.
  Depending on the application's energy-delay trade-off, the station may wake up before, on, or after the predicted time.
  }
  \label{RSgraph}
\end{figure}

We consider three variants of EAPS based on prediction error.
To justify these variations, we first present the distribution of prediction errors in Figure \ref{RSgraph} for voice and background \gls{AC}s.
Based on distribution for each AC, the station can either choose to wake up at (i) $\delta^{\prime} - 2\sigma$, (ii) $\delta^{\prime} + 2\sigma$, or (iii) $\delta^{\prime}$.
We call these cases \gls{EAPS-E}, \gls{EAPS-L}, and \gls{EAPS-M}.
Intuitively, \gls{EAPS-E} reduces delay with a higher energy consumption, \gls{EAPS-L} reduces energy with a longer delay, and \gls{EAPS-M} establishes a trade-off between energy and delay.
Please note that \gls{EAPS-E} is only applicable if $\delta^{\prime} - 2\sigma > 0$.


\subsection{Results}

\begin{figure*}[t]
\centering
  \includegraphics[width=0.98\textwidth]{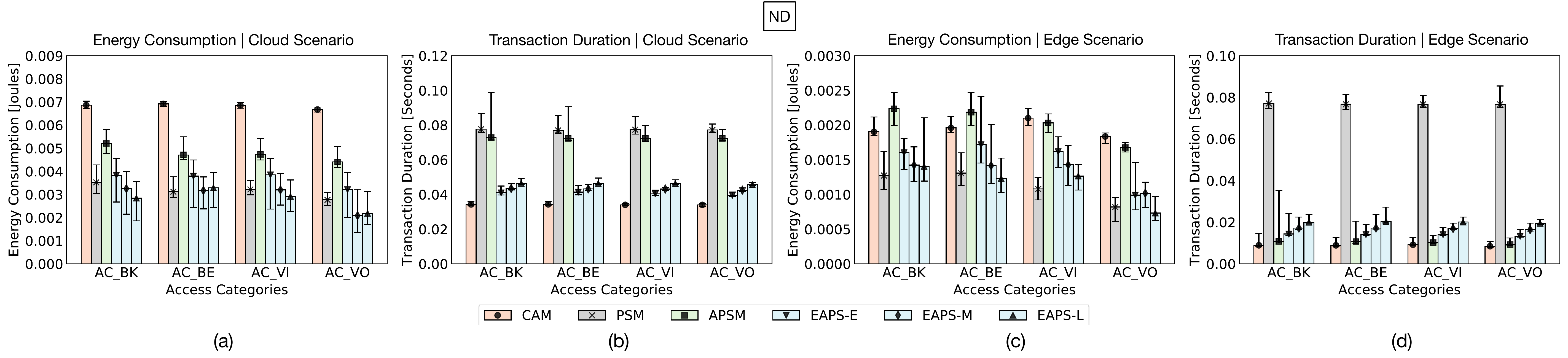}
  \caption{Performance comparison of EAPS with 802.11 power saving mechanisms in \textit{ND} conditions for all ACs. 
  (a) and (b) show the average \textit{per-transaction} energy and duration in cloud scenario, respectively.  
  (c) and (d) show the average \textit{per-transaction} energy and duration in edge scenario, respectively. 
  }
  \label{ND_em_eval}
\end{figure*}

\begin{figure*}[t]
\centering
  \includegraphics[width=0.98\textwidth]{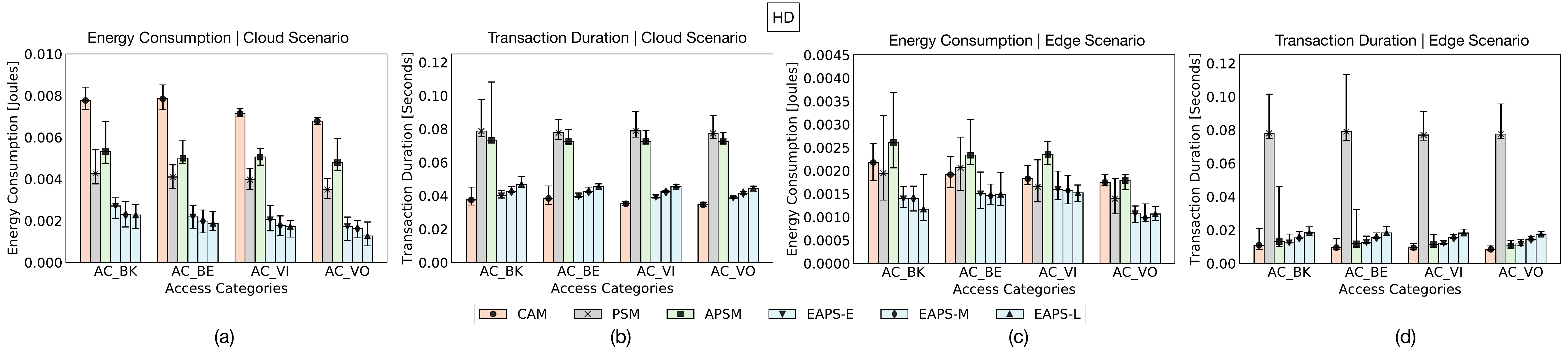}
  \caption{Performance comparison of EAPS with 802.11 power saving mechanisms in \textit{HD} conditions for all ACs. 
  (a) and (b) show the average \textit{per-transaction} energy and duration in cloud scenario, respectively.  
  (c) and (d) show the average \textit{per-transaction} energy and duration in edge scenario, respectively. 
  }
  \label{HD_em_eval}
\end{figure*}

Figures \ref{ND_em_eval} and \ref{HD_em_eval} illustrate the average energy consumption and duration of transactions when the station is communicating with edge and cloud computing platforms under ND and HD conditions, respectively.

%
%

In the cloud computing scenario, CAM and EAPS incur an average round trip delay of 35 ms and 42 ms, respectively, while EAPS consumes 63\% less energy.
This is because EAPS conserves energy expenditure by switching to sleep mode and waking up right before the packet is ready for transmission at AP. 
In contrast, CAM needs to stay in awake mode until the response is received.
Reduction in energy consumption of EAPS compared to CAM reduces to 30\% in edge environment due to the shorter duration spent in awake mode to receive the downlink packet.

With PSM, the station immediately transitions to sleep mode after transmitting each uplink packet.
While this results in less energy consumption compared to CAM, transactions suffer about 55 ms higher delay on average because the earliest opportunity for downlink packet delivery is after the next beacon instance.
The transaction duration of EAPS is 62\% lower compared to PSM on average across all the ACs.
With APSM, the station remains in idle (wait) state for 10 ms after transmission or reception.
This is beneficial only in specific scenarios.
For example, in the edge scenario, the station receives its downlink packet within the tail time (similar to CAM).
However, when the round trip delay is more than 10 ms, the station will have to wake up again to retrieve the downlink packet after the next beacon announcement, thereby resulting in both long delay and waste of more energy compared to PSM.
On average, for both edge and cloud scenarios, the energy consumption of APSM is 30\% higher compared to PSM.
In contrast, the energy consumption of EAPS is 20\% and 43\% lower than PSM and APSM, respectively.
Also, the transaction duration of PSM, APSM, and EAPS are 77 ms, 10 ms, and 12 ms in edge computing scenario, and 77 ms, 72 ms, and 42 ms in cloud computing scenario.


With EAPS, the station has the freedom to choose between \gls{EAPS-E}, \gls{EAPS-M}, or \gls{EAPS-L}, according to application requirements.
We observed that with \gls{EAPS-E}, the station suffers from slightly higher energy consumption because it wakes up early, waits for the packet to be received from the AP, and then transitions into sleep mode.
In the case of \gls{EAPS-L}, since the station wakes up $2\times\sigma$ after the predicted delay, the downlink packet will be buffered at the AP.
However, the station can immediately transition to sleep mode once the packet is received.
Thus, energy consumption of \gls{EAPS-L} is 14\% lesser compared to \gls{EAPS-E}, whereas, the transaction duration of \gls{EAPS-L} is 18\% higher than \gls{EAPS-E}. 
\gls{EAPS-M} balances the trade off between energy consumption and transaction duration.

\section{Related Work}
\label{related_work}
Peck et al. \cite{peck2015practical} propose PSM with adaptive wake-up (PSM-AW), which includes a metric called \textit{PSM penalty} to enable the user to establish their desired energy-delay trade-off.
The authors define server delay as the total delay between sending a request to a server and receiving a reply.
Based on RTT variations, the sleep duration is dynamically adjusted to satisfy the desired trade-off.
Also, the size of the history window of server variations is dynamically tuned based on the range of server delay variations.
Compared to our presented work, PSM-AW \cite{peck2015practical} assumes the significant component of the uplink-downlink delay is communication time with the server, thereby ignoring the variable, long impact of downlink wireless communication delay.
Also, RTT sampling and averaging are performed by stations.
Therefore, delay estimation is directly affected by station-server communication, and estimation accuracy drops as the interval between transactions increases.
In contrast, our work does not impose overhead on stations, and once a model is trained, it does not rely on ongoing communication to compute sleep schedules.
Jang et al. \cite{jang2016adaptive} proposed an adaptive tail time adjustment mechanism by relying on inter-packet arrival delays.
A moving average scheme is used to predict inter-packet arrival delay when a burst of packets arrives at a station.
The station stays in the awake mode if the next packet arrival time is before the tail time expiry.
If packet delivery is after the tail time expiry, the station might extend its tail time based on the outcome of an energy-delay trade-off model.
In contrast to our solution, neither \cite{peck2015practical} nor \cite{jang2016adaptive} considers the impact of buffering and channel access delay as variable and essential components of downlink delivery delay.
Furthermore, the effectiveness of these approaches highly depends on the burst length and variability of end-to-end delays.
Specifically, the moving average scheme employed in \cite{jang2016adaptive} would not be effective in IoT scenarios where most of the bursts are short-lived.
Sui et al. \cite{sui2016characterizing} propose WiFiSeer, a centralized decision-making system to help stations choose the \gls{AP} providing the shortest delay.
WiFiSeer works in two phases: During the learning phase, a set of parameters (such as RSSI, RTT) are pulled every minute from all \gls{AP}s using SNMP. 
Then a random forest model is trained to generate a two-class learning model for classifying \gls{AP}s into high latency and low latency.
A user agent installed on smartphones communicates with the controller and associates the station with the recommended \gls{AP}.
WiFiSeer is complementary to our solution to further reduce station-\gls{AP} delay.

Jang et al. \cite{jang2011snooze} study the overhead of radio switching and show that stations can achieve significant energy saving during inter-frame delays while the \gls{AP} is communicating with other stations.
The proposed \gls{AP}-driven approach, called Snooze, utilizes the global knowledge of the \gls{AP} to schedule sleep and awake duration of each station based on inter-packet delays and traffic load of the station.
To distribute the schedule, the \gls{AP} needs to exchange control messages with stations. 
Compared to Snooze, our work considers the sensing-actuation pattern of IoT applications and reduces the idle listening time between sensing and actuation.
In addition, our work takes into account the impact of interference by measuring airtime utilization perceived by the \gls{AP}.
Also, Snooze does not benefit from \gls{APSD}.
Sheth and Dezfouli \cite{sheth2019enhancing} propose \textit{Wiotap}, an \gls{AP}-based, layer-3 scheduling mechanism for IoT stations that employ \gls{APSM}.
Their proposed mechanism uses an \gls{EDF} scheduling strategy to maximize the chance of packet delivery before tail time expiry.
Rozner et al. \cite{rozner2010napman} propose NAPman, which prioritizes the delivery of \gls{PSM} traffic as long as other stations are not affected.
Tozlu et al. \cite{tozlu2012wi} show that increasing \gls{AP} load has a higher impact on packet loss and RTT compared to out-of-band interference.
Pei et al. \cite{pei2016wifi} argue that 802.11 link delay comprises more than 60\% of round trip delay.
They also show that more than 50\% (10\%) of packets experience more than 20 ms (100 ms) latency by station-\gls{AP} links.
For TCP traffic, the authors proposed an approach to measure the delay of wired latency as well as uplink and downlink channel access delay.
Using the Kendall correlation, they also show that airtime utilization, RSSI, and retry rate are the top three factors affecting station-\gls{AP} delay.
Their study also shows that the physical rate offers the highest relative information gain to predict latency.
This is because of physical rate changes based on RSSI, frame loss, and channel quality.
Then, a decision tree model is used to tune the parameters of \gls{AP}s and reduce the overall delay experienced by stations.
This is in contrast to our work, which offers per-station, fine-granularity sleep scheduling.
Primarily designed for VoIP traffic, Liu et al. \cite{liu2014energy} propose a mechanism to reduce contention among stations waking up using \gls{APSD} to retrieve packets from the \gls{AP}.
After receiving a burst of voice data, the station measures the tolerable deadline of incoming packets and informs the \gls{AP} about its wake up time before switching to sleep mode.
The wake-up instance is approved only if the \gls{AP} will be idle when the station will wake up.




\section{Conclusion}
\label{conclusion}
In this paper, we presented the design and implementation of a predictive scheduling mechanism, named EAPS, to enable IoT stations to transition to sleep mode and wake up when the downlink packet is expected to be delivered.
The proposed solution benefits from edge computing, meaning the computations of sleep scheduling are performed at the network edge and by the \gls{AP}.
We presented an \gls{AP} architecture capable of collecting queues status, channel condition, and packet transmission and reception instances.
Once the \gls{AP} receives an uplink packet, a machine learning model is used to compute the sleep delay, and the station is informed about its schedule using a high-priority data plane.
Using empirical evaluations, we confirm the significant enhancement of EAPS in terms of energy efficiency and transaction delay.

EAPS can be considered as an enhancement of the \gls{APSD} and \gls{TWT} (introduced in 802.11ax).
For example, with the next generation of IoT stations that support \gls{TWT}, they can set up their wake up time based on the sleep schedule computed by the \gls{AP}.
By protecting IoT stations against the effect of concurrent traffic and interference, EAPS is a particularly useful mechanism in scenarios where both regular and IoT stations exist.
For example, EAPS can reduce the energy cost of households and reduce the impact of IoT on global ICT footprint.

To extend this work, we plan to integrate EAPS with a software-defined architecture and enable the stations to receive their downlink packet from the \gls{AP} offering the lowest delay.
We also plan to incorporate a predictive buffering mechanism to provide delay bound guarantees for mission-critical applications.
EAPS can also significantly benefit from wake-up radio technology:
After sending the uplink packet(s), the station can immediately transition into sleep mode and use its secondary low-power radio to receive the schedule message.
When it is time to receive the downlink packet, the secondary radio can first communicate with the \gls{AP} to verify if the downlink packet is ready to be sent.

\balance


%





\ifCLASSOPTIONcaptionsoff
  \newpage
\fi



%

\bibliographystyle{IEEEtran}

\bibliography{references}

%







\end{document}

%% file: list_notations.tex
\newcommand{\Scheduler}{\texttt{Scheduler}}
\newcommand{\Collector}{\texttt{Collector}}
\newcommand{\Sniffer}{\texttt{Sniffer}}
\newcommand{\NetMon}{\texttt{NetMon}}
\newcommand{\ChUMon}{\texttt{ChUMon}}
\newcommand{\NetQMon}{\texttt{NetQMon}}
\newcommand{\MACQMon}{\texttt{MACQMon}}
\newcommand{\netfilter}{\texttt{netfilter}}

%% file: list_abbreviations.tex
\newacronym{IoT}{IoT}{Internet of Things}

\newacronym{SCU}{SCU}{Santa Clara University}
\newacronym{UTM}{UTM}{University Technology Malaysia}
\newacronym{SES}{SES}{Summer Engineering Seminar}

\newacronym{RTOS}{RTOS}{Real-Time Operating System}

\newacronym{MQTT}{MQTT}{Message Queue Telemetry Transport}

\newacronym{API}{API}{Application Programming Interface}

\newacronym{WSN}{WSN}{Wireless Sensor Networks}

\newacronym{EDF}{EDF}{Earliest Deadline First}

\newacronym{MAE}{MAE}{Mean Absolute Error}

\newacronym{RFR}{RFR}{Random Forest Regressor}
\newacronym{GBR}{GBR}{Gradient Boosting Regressor}
\newacronym{ABR}{ABR}{Adaptive Boosting Regressor}
\newacronym{ETR}{ETR}{Extra Trees Regressor}
\newacronym{LSTM}{LSTM}{Long Short-Term Memory}
\newacronym{HBR}{HBR}{Histogram-Based Gradient Boosting Regressor}
\newacronym{NN}{NN}{Neural Networks}
\newacronym{RNN}{RNN}{Recurrent Neural Networks}

\newacronym{H2H}{H2H}{Humand to Human}
\newacronym{M2H}{M2H}{Machine to Human}
\newacronym{M2M}{M2M}{Machine to Machine}

\newacronym{MAC}{MAC}{Medium Access Control}
\newacronym{CW}{CW}{Contention Window}

\newacronym{RAT}{RAT}{Radio Access Technology}

\newacronym{CAGR}{CAGR}{Compound Annual Growth Rate}

\newacronym{RSSI}{RSSI}{Received Signal Strength Indicator}
\newacronym{SNR}{SNR}{Singal-to-Noise Ratio}

\newacronym{BLE}{BLE}{Bluetooth Low Energy}

\newacronym{NFC}{NFC}{Near Field Communication}

\newacronym{WEP}{WEP}{Wired Equivalent Privacy }
\newacronym{WPA}{WPA}{Wireless Protected Access}
\newacronym{PMK}{PMK}{Pairwise Master Key}
\newacronym{PTK}{PTK}{Pairwise Transition Keys}
\newacronym{GTK}{GTK}{Group Temporal Key}
\newacronym{EAP}{EAP}{Extensible Authentication Protocol}
\newacronym{RRM}{RRM}{Radio Resource Management}
\newacronym{BSSTM}{BSSTM}{BSS Transition Management}
\newacronym{BSS}{BSS}{Basic Service Set}
\newacronym{SSID}{SSID}{Service Set Identifier}
\newacronym{BSSID}{BSSID}{BSS Identifier}
\newacronym{CSA}{CSA}{Channel Switching Announcement}
\newacronym{DCF}{DCF}{Distributed Coordination Function}
\newacronym{EDCF}{EDCF}{Enhanced Distributed Coordination Function}
\newacronym{AIFS}{AIFS}{Arbitration Inter-Frame Space}

\newacronym{QoS}{QoS}{Quality of Service}
\newacronym{ToS}{ToS}{Type of Service}

\newacronym{TSF}{TSF}{Time Synchronization Function}
\newacronym{AC}{AC}{Access Category}
\newacronym{PSM}{PSM}{Power Save Mode}
\newacronym{PSP}{PSP}{Power Save Poll}
\newacronym{PSP-N}{PSP-N}{Power Save Poll with Null}
\newacronym{CAM}{CAM}{Continuously Active Mode}
\newacronym{AID}{AID}{Association ID}
\newacronym{APSM}{APSM}{Adaptive PSM}
\newacronym{APSD}{APSD}{Automatic Power Save Delivery}
\newacronym{U-APSD}{U-APSD}{Unscheduled-Automatic Power Save Delivery}
\newacronym{S-APSD}{S-APSD}{Scheduled-Automatic Power Save Delivery}
\newacronym{EOSP}{EOSP}{End of Service Period}
\newacronym{SP}{SP}{Service Period}
\newacronym{TXOP}{TXOP}{Transmit Opportunity}
\newacronym{BI}{BI}{Beacon Interval}

\newacronym{DBS}{DBS}{Disassociation-Based Steering}

\newacronym{SNMP}{SNMP}{Simple Network Management Protocol}

\newacronym{ODL}{ODL}{OpenDayLight} 
\newacronym{OVS}{OVS}{Open vSwitch}
\newacronym{OVSDB}{OVSDB}{Open vSwitch Database}
\newacronym{SDN}{SDN}{Software Defined Networking}

\newacronym{VM}{VM}{Virtual Machine}

\newacronym{NF}{NF}{Network Function}
\newacronym{VNFM}{VNFM}{VNF Manager}
\newacronym{VNF}{VNF}{Virtual Network Function}
\newacronym{NNF}{NNF}{Native Network Function}
\newacronym{NFVO}{NFVO}{NFV Orchestrator}
\newacronym{VIM}{VIM}{Virtualized Infrastructure Manager}

\newacronym{PCP}{PCP}{Priority Code Point}

\newacronym{NAT}{NAT}{Network Address Translation}

\newacronym{NUMA}{NUMA}{Non-Uniform Memory Access}

\newacronym{CPE}{CPE}{Customer Premise Equipment}

\newacronym{DSCP}{DSCP}{Differentiated Services Code Point}

\newacronym{vSG}{vSG}{Virtual Subscriber Gateway}

\newacronym{RGW}{RGW}{Residential Gateway}
\newacronym{vRGW}{vRGW}{Virtual RGW}

\newacronym{VF}{VF}{Virtual Function}
\newacronym{PF}{PF}{Physical Function}

\newacronym{ARP}{ARP}{Address Resolution Protocol}

\newacronym{TIM}{TIM}{Traffic Indication Map}
\newacronym{DTIM}{DTIM}{Delivery Traffic Indication Message}

\newglossaryentry{AP}
{
  name={AP},
  description={Access Point},
  first={\glsentrydesc{AP} (\glsentrytext{AP})},
  plural={APs},
  firstplural={\glsentrydesc{AP}s (\glsentryplural{AP})}
}

\newacronym{TWT}{TWT}{Target Wake-up Time}
\newacronym{OPS}{OPS}{Opportunistic Power Save}
\newacronym{PCF}{PCF}{Point Coordination Function}
\newacronym{PIFS}{PIFS}{PCP Inter Frame Space}
\newacronym{DIFS}{DIFS}{Distributed Inter Frame Space}
\newacronym{BSR}{BSR}{Buffer Status Report}
\newacronym{OCW}{OCW}{OFDM Contention Window}
\newacronym{OFDMA}{OFDMA}{Orthogonal Frequency-Division Multiple Access}
\newacronym{OBO}{OBO}{OFDMA Back-off}
\newacronym{RU}{RU}{Resource Unit}

\newacronym{QTP}{QTP}{Quite Time Period}
\newacronym{AU}{AU}{Airtime Utilization}

\newacronym{LVAP}{LVAP}{Light Virtual AP}
\newacronym{EDCA}{EDCA}{Enhanced Distributed Channel Access}

\newacronym{HCCA}{HCCA}{Hybrid Controlled Channel Access}

\newacronym{WOL}{WOL}{Wake-on-Lan}

\newacronym{DC}{DC}{Datacenter}
\newacronym{CO}{CO}{Central Office}

\newacronym{CORD}{CORD}{Central Office Re-architected as a Datacenter}
\newacronym{MEC}{MEC}{Mobile Edge Computing}
\newacronym{ASP}{ASP}{Application Service Provider}
\newacronym{ES}{ES}{Edge Service}
\newacronym{NSP}{NSP}{Network Service Provider}

\newacronym{NV}{NV}{Network Virtualization}
\newacronym{NFV}{NFV}{Network Function Virtualization}
\newacronym{NFVI}{NFV}{NFV Infrastructure}
\newacronym{MANO}{MANO}{Management and Orchestration}

\newacronym{NIC}{NIC}{Network Interface Card}
\newacronym{vNIC}{vNIC}{virtual NIC}
\newacronym{pNIC}{pNIC}{physical NIC}

\newacronym{TCAM}{TCAM}{Tenary Content Addressable Memory}
\newacronym{RSS}{RSS}{Receive Side Scaling}

\newacronym{DPDK}{DPDK}{Data Plane Development Kit}
\newacronym{UIO}{UIO}{User-space I/O}

\newacronym{EMC}{EMC}{Exact Match Cache}

\newacronym{SR-IOV}{SR-IOV}{Single Root-Input/Output Scaling}
\newacronym{PMD}{PMD}{Poll Mode Driver}

\newacronym{TSS}{TSS}{Tuple Space Search}
\newacronym{dpcls}{dpcls}{data path classifier}

\newacronym{FDK}{FDK}{Fog Development Kit}
\newacronym{RAA}{RAA}{Resource Allocation Algorithm}

\newacronym{TSN}{TSN}{Time Sensitive Networking}
\newacronym{AFDX}{AFDX}{avionics full-duplex switched Ethernet}
\newacronym{CAN}{CAN}{Controller Area Network}

\newacronym{OCI}{OCI}{Open Carrier Interface }

\newacronym{CBR}{CBR}{constant bit rate}

\newacronym{EAPS-E}{EAPS-E}{EAPS with Early wake-up}
\newacronym{EAPS-L}{EAPS-L}{EAPS with Late wake-up}
\newacronym{EAPS-M}{EAPS-M}{EAPS with Mid wake-up}

\newacronym{ECDF}{ECDF}{Empirical Cumulative Distribution Function}